%% file: ANA-STDM-2018-14-PAPER.tex
\pdfoutput=1
\documentclass[PAPER,atlasdraft=False,cernpreprint,texlive=2016,UKenglish,texmf]{atlasdoc}
 
\usepackage{atlaspackage}
\usepackage{atlasbiblatex}
 
\usepackage{atlascontribute}
 
\usepackage[jetetmiss,hepprocess]{atlasphysics}

\graphicspath{{logos/}{figures/}}
 
\usepackage{ANA-STDM-2018-14-PAPER-defs}

\AtlasTitle{Measurement of the transverse momentum distribution of Drell--Yan lepton pairs in proton--proton collisions at $\sqrt{s}=13\,$TeV with the ATLAS detector}
 
\AtlasJournalRef{Eur. Phys. J. C 80 (2020) 616}
\AtlasDOI{10.1140/epjc/s10052-020-8001-z}
 
\AtlasAbstract{
This paper describes precision measurements of the transverse momentum $p_\mathrm{T}^{\ell\ell}$ ($\ell=e,\mu$) and of the angular variable $\phi^{*}_{\eta}$ distributions of Drell--Yan lepton pairs in a mass range of 66--116$\,$GeV. The analysis uses data from 36.1$\,$fb$^{-1}$ of proton--proton collisions at a centre-of-mass energy of $\sqrt{s}=13\,$TeV collected by the ATLAS experiment at the LHC in 2015 and 2016. Measurements in electron-pair and muon-pair final states are performed in the same fiducial volumes, corrected for detector effects, and combined. Compared to previous measurements in proton--proton collisions at $\sqrt{s}=7$ and $8\,$TeV, these new measurements probe perturbative QCD at a higher centre-of-mass energy with a different composition of initial states. They reach a precision of 0.2$\%$ for the normalized spectra at low values of $p_\mathrm{T}^{\ell\ell}$. The data are compared with different QCD predictions, where it is found that predictions based on resummation approaches can describe the full spectrum within uncertainties.}

\author{The ATLAS Collaboration}

\PreprintIdNumber{CERN-EP-2019-223}

\AtlasRefCode{ANA-STDM-2018-14}
 
\AtlasNote{ANA-STDM-2018-14-PAPER}
\hypersetup{pdftitle={ATLAS document},pdfauthor={The ATLAS Collaboration}}
 
\begin{document}
 
\maketitle
 
\tableofcontents

% The next lines are included from the .//Sec_1_Introduction.tex input file
\section{Introduction}
 
In high-energy hadron--hadron collisions, the vector bosons \Wboson and \Zg are produced via quark--antiquark annihilation~\cite{Drell:1970wh}, and can be observed with very small backgrounds by using their leptonic decay modes. The vector bosons have non-zero momentum transverse to the beam direction due to the emission of quarks and gluons from the initial-state partons as well as to the intrinsic transverse momentum of the initial-state partons in the proton. Phenomenologically, the spectrum at low transverse momentum of the $Z$ boson, \Zpt{}, reconstructed through the decay into a pair of charged leptons, can be described using soft-gluon resummation~\cite{Collins:1984kg,Balazs:1997xd, Becher:2010tm, Catani:2015vma, Ebert:2016gcn, Bizon:2017rah} and non-perturbative models to account for the intrinsic transverse momentum of partons. At high \Zpt{} the spectrum can be calculated by fixed-order perturbative quantum chromodynamics (QCD) predictions~\cite{Melnikov:2006kv,Li:2012wna,Catani:2009sm,Boughezal:2015ded,Ridder:2015dxa}, and next-to-leading-order electroweak (NLO EW) effects are expected to be important~\cite{Denner:2011vu,Kallweit:2015dum,Lindert:2017olm}. Parton-shower models~\cite{Bellm:2015jjp,Sjostrand:2014zea,Bothmann:2019yzt} or resummation may be matched to fixed-order calculations to describe the full spectrum.
 
A precise measurement of the \Zpt{} spectrum provides an important input to the background prediction in searches for beyond the Standard Model (SM) processes, e.g.\ in the monojet signature~\cite{EXOT-2016-27}, as well as to SM precision measurements. In particular, the measurement of the mass of the \Wboson{} boson~\cite{STDM-2014-18} relies on the measurement of the \Zpt{} distribution to constrain the transverse momentum spectrum of the \Wboson{} boson, \pTW{}, since a direct measurement of the transverse momentum distribution of \Wboson{} bosons is experimentally challenging \cite{STDM-2011-15}. The \Zpt{} spectrum was measured previously in proton--proton ($pp$) collisions at the Large Hadron Collider (LHC) by the ATLAS Collaboration at centre-of-mass energies of $\sqrt{s} = 7$~\TeV{} and $8$~\TeV~\cite{STDM-2012-23, STDM-2014-12}, including several mass regions near and away from the \Zboson{}-boson resonance. Related measurements were also made by the CMS~\cite{CMS-SMP-17-010, CMS-SMP-17-002, CMS-SMP-14-012, CMS-SMP-13-013, CMS-EWK-10-010} and the LHCb~\cite{Aaij:2015gna,AbellanBeteta:2016ugk,Aaij:2016mgv} collaborations at the LHC and by the CDF~\cite{Aaltonen:2012fi} and D0~\cite{Abazov:2010kn,Abazov:2010mk} collaborations in $p\bar{p}$ collisions at the TeVatron.
 
Compared to measurements at lower $\sqrt{s}$, \Zboson{}-boson
production at 13~\TeV\ is characterized by a smaller parton momentum
fraction of the colliding protons, leading to a different flavour
composition and a larger phase space for hard QCD radiation. A precise
measurement will test this energy dependence and play an important role in
future studies of the \Wboson{}-boson mass using the 13~\TeV{} data.
 
The granularity of the measurement in the low-\Zpt\ domain is limited by the lepton
momentum resolution. To overcome this limitation, the
$\phi_\eta^*$ observable was introduced~\cite{Banfi:2010cf} as an
alternative probe of \Zpt{}. It is defined as
 
\begin{equation*}
\phi_\eta^* = \tan\left(\frac{\pi-\Delta\phi}{2}\right) \times \sin(\theta^*_\eta)\,,
\end{equation*}
 
where $\Delta \phi$ is the azimuthal angle in radians between the two leptons. The angle $\theta_\eta^*$ is a measure of the scattering angle of the leptons relative to the proton beam direction in the rest frame of the dilepton system and is defined by $\cos(\theta_\eta^*) = \tanh[(\eta^- - \eta^+)/2]$, where $\eta^-$ and $\eta^+$ are the pseudorapidities\footnote{ ATLAS uses a right-handed coordinate system with its origin at the nominal interaction point (IP) in the centre of the detector and the $z$-axis along the beam pipe.  The $x$-axis points from the IP to the centre of the LHC ring, and the $y$-axis
points upwards.  Cylindrical coordinates $(r,\phi)$ are used in the
transverse plane, $\phi$ being the azimuthal angle around the
$z$-axis.  The pseudorapidity is defined in terms of the polar angle
$\theta$ as $\eta = -\ln \tan(\theta/2)$. Angular distance is
measured in units of $\Delta R = \sqrt{(\Delta\eta)^{2} +
(\Delta\phi)^{2}}$.  } of the negatively and positively charged lepton, respectively. Therefore, \phiStar{} depends exclusively on the directions of the two leptons, which are measured more precisely than their momenta.
 
In this paper, measurements of the \Zpt{} and the \phiStar{} spectra
are presented using $pp$ collision data at $\sqs=13$~\TeV\ collected in
2015 and 2016 with the ATLAS detector, corresponding to an integrated
luminosity of \lumi. Both the dielectron and dimuon final states \Zgll ($\ell = e$ or $\mu$) are analysed in a dilepton mass window
of $\Mll = 66$--$116$~\GeV. The measurement is performed in a fiducial
phase space that is close to the detector acceptance for leptons in
transverse momentum $\pt^\ell$ and pseudorapidity
$\eta_\ell$.
 
% End of text imported from the .//Sec_1_Introduction.tex input file
% The next lines are included from the .//Sec_2_ATLAS.tex input file
\section{The ATLAS detector} 
The ATLAS experiment uses a multipurpose detector~\cite{PERF-2007-01, PIX-2018-001, ATLAS-TDR-2010-19} with a cylindrical geometry and almost $4\pi$ coverage in solid angle. The collision point is surrounded by tracking detectors, collectively referred to as the inner detector (ID), followed by a superconducting solenoid providing a 2~T axial magnetic field, a calorimeter system and a muon spectrometer.
The ID provides precise measurements of charged-particle tracks in the pseudorapidity range $|\eta|<2.5$.
It consists of three subdetectors arranged in a coaxial geometry around the beam axis: a silicon pixel detector,
a silicon microstrip detector and a transition radiation tracker.
 
The electromagnetic calorimeter covers the region $|\eta|<3.2$ and is based on a high-granularity, lead/liquid- argon (LAr) sampling technology.
The hadronic calorimeter uses a steel/scintillator-tile detector in the region $|\eta|<1.7$ and a copper/LAr detector in the region $1.5<|\eta|<3.2$.
The forward calorimeter (FCAL)
covers the range $3.2< |\eta| < 4.9$ and also uses LAr as the active
material and copper or tungsten absorbers for the EM and hadronic sections, respectively.
 
The muon spectrometer (MS) consists of separate trigger and high-precision tracking chambers to measure the deflection of muons in a magnetic field generated by three large superconducting toroids arranged with an eightfold azimuthal coil symmetry around the calorimeters.
The high-precision chambers cover a range of $|\eta|<2.7$.
The muon trigger system covers the range $|\eta|<2.4$ with resistive-plate chambers in the barrel and thin-gap chambers in the endcap regions.
 
A two-level trigger system is used to select events in real time~\cite{TRIG-2016-01}. It consists of a hardware-based first-level trigger and a software-based high-level trigger. The latter employs algorithms similar to those used offline and is used to identify electrons and muons.
% End of text imported from the .//Sec_2_ATLAS.tex input file
% The next lines are included from the .//Sec_3_AnalysisMethodology.tex input file
\section{Analysis methodology} 
 
\subsection{Description of the measurements\label{sec:desc}}
 
The \Zboson-boson differential cross-sections are measured as a function of \Zpt{} and \phiStar{} separately for the dielectron and dimuon decay channels. Only small background contributions are expected. The results are reported within a fiducial phase space chosen to be close to the experimental acceptance defined by the lepton transverse momenta $\pt^\ell > 27$~\GeV, the absolute lepton pseudorapidity $|\eta_\ell| < 2.5$ and the dilepton invariant mass $\Mll = 66$--$116$~\GeV.
 
The lepton kinematics can be described at different levels regarding the effects of final-state photon radiation (QED FSR). Cross-sections at \textit{Born} level employ the lepton kinematics before QED FSR, while the \textit{bare} level is defined by leptons after emission of QED FSR. A \textit{dressed} lepton is defined by combining the bare four-momenta of each lepton with that of QED FSR photons radiated from the lepton within a cone of size $\Delta R = 0.1$ around the lepton. The results in this paper are reported at the dressed and Born levels.
 
The differential cross-sections in \Zpt{} and \phiStar{} are measured and their normalized spectra derived. The total systematic uncertainty of the latter is significantly reduced due to large correlations in many sources of uncertainty between the measurement bins.

\subsection{Simulated event samples}
\label{sec:datamc}
 
Events simulated by Monte Carlo (MC) generators are used to predict the detector response to the signal process in order to correct the data for detector inefficiencies and resolution as well as to estimate most of the background from processes other than $\Zg \to\ell\ell$ in the selected data sample.
 
The \Zgll signal process was generated with the \textsc{Powheg-Box V1} MC event generator~\cite{Nason:2004rx,Frixione:2007vw,Alioli:2010xd,Alioli:2008gx} at next-to-leading order in $\alphas$ interfaced to \PYTHIAV{8.186}~\cite{Sjostrand:2014zea} for the modelling of the parton shower, hadronization, and underlying
event, with parameters set according to the AZNLO
tune~\cite{STDM-2012-23}. The CT10 (NLO) set of parton distribution functions (PDF) ~\cite{Lai:2010vv} was used
for the hard-scattering processes, whereas the CTEQ6L1 PDF
set~\cite{Pumplin:2002vw} was used for the parton shower. The effect of final-state photon radiation was simulated with Photos++~v3.52~\cite{Golonka:2005pn,Davidson:2010ew}. The \textsc{EvtGen}~v1.2.0
program~\cite{EvtGen} was used to decay bottom and charm hadrons.
 
\POWHEG{}+\PYTHIA{}8 was also used to simulate the majority of the background processes considered. The \Ztt and the diboson processes $WW$, $WZ$ and $ZZ$~\cite{Nason:2013ydw} (requiring $m_{\ell\ell}>4$~\GeV\ for any pair of same-flavour opposite-charge leptons) used the same tune and PDF as the signal process. The \ttbar and single-top-quark~\cite{Frixione:2007nw,Re:2010bp} backgrounds to the dielectron channel were simulated with \POWHEG{}+\PYTHIA{}6~\cite{Sjostrand:2006za} with the P2012 tune~\cite{Skands:2010ak} and CT10 PDF, while for the dimuon channel \POWHEG{}+\PYTHIA{}8 with the A14 tune~\cite{ATL-PHYS-PUB-2014-021} and the NNPDF3.0 PDF~\cite{Ball:2014uwa} was used. It was found that the prediction of the \ttbar background is in very good agreement for both generators.
The photon-induced background $\gamma\gamma\rightarrow \ell\ell$ was generated with \PYTHIA{}8 using the NNPDF2.3 QED PDF~\cite{Ball:2013hta}.
 
The effect of multiple interactions in the same and neighbouring bunch crossings (pile-up) was modelled by overlaying the hard-scattering event with simulated minimum-bias
events generated with \PYTHIAV{8.186} using the MSTW2008LO set of PDFs~\cite{Martin:2009iq} and the A2 tune~\cite{ATL-PHYS-PUB-2012-003}.
The simulated event samples were reweighted to describe the distribution of the number of pile-up interactions in the data, and further reweighted such that the distribution of the longitudinal position of the primary $pp$ collision vertex matches that in data. The primary vertex is defined as the vertex with at least two reconstructed tracks with $\pt>0.4$ GeV and with the highest sum of squared transverse momenta of associated tracks. The \GEANT{}4 program was used to simulate the passage of particles through the ATLAS detector~\cite{Agostinelli:2002hh,SOFT-2010-01}.
The simulated events are reconstructed with the same analysis procedure as the data.
The reconstruction, trigger and isolation efficiencies as well as lepton momentum scale and resolution in the MC simulation are corrected to match those determined in data~\cite{PERF-2015-10,PERF-2017-01,PERF-2017-03}.

\subsection{Event selection}

Candidate Z $\rightarrow ee$ events are triggered requiring
at least one identified electron with $\pt > 24$~\GeV\ in 2015
and $\pt > 26$~\GeV\ in 2016 data~\cite{TRIG-2018-05}. In addition to the increased $\pt$ threshold, the electron also has to
satisfy isolation criteria in the 2016 data.
Candidate \Zmm{} events were recorded with triggers that require at least one isolated muon with $\pt > 20$~\GeV\ in 2015 and $\pt > 26$~\GeV\ in 2016 data.
 
Electron candidates are reconstructed from clusters of energy in the electromagnetic calorimeter matched to ID tracks~\cite{PERF-2017-01}.
They are required to have $\pt > 27$~\GeV\ and $|\eta|<2.47$ (excluding the transition regions between the barrel and the endcap electromagnetic calorimeters, $1.37<|\eta|<1.52$). Electron candidates are required to pass the `medium` identification requirement, and are also required to be isolated according to the `gradient' isolation criterion~\cite{PERF-2017-01}.
 
Muon candidates are reconstructed by combining tracks reconstructed in the inner detector with tracks reconstructed in the MS~\cite{PERF-2015-10}. They are required to have $\pt > 27$~\GeV\ and $|\eta|<2.5$ and satisfy identification criteria corresponding to the `medium' working point~\cite{PERF-2015-10}. Track quality requirements are imposed to suppress backgrounds, and the muon candidates are required to be isolated according to the `gradient' isolation criterion~\cite{PERF-2015-10}, which is \pT- and $\eta$-dependent and based on the calorimeter and track information.
 
Electron and muon candidates are required to originate from the primary vertex. Thus, the significance of the track’s transverse impact parameter calculated relative to the beam line, $|d_0/\sigma_{d_{0}}|$, must be smaller than 3.0 for muons and less than 5.0 for electrons. Furthermore, the longitudinal impact parameter, $z_0$ (the difference between the $z$-coordinate of the point on the track at which $d_0$ is defined and the longitudinal position of the primary vertex), is required to satisfy $|z_0 \cdot \sin(\theta)| <$ 0.5 mm.
 
Events are required to contain exactly two same-flavour leptons passing the lepton selection. The two leptons must be of opposite electric charge and their invariant mass must satisfy $66 < \Mll < 116$~\GeV. No additional veto on the presence of leptons of different flavour is applied. Table~\ref{tab:DataZleptons} shows the number of events satisfying the above selection criteria in the electron channel and the muon channel. Also given are the estimated contributions from the background sources described below in Section~\ref{sec:backgrounds}.
 
\begin{table}[thb]
\caption{\label{tab:DataZleptons}Selected signal candidate events in data for both decay channels as well as the expected background contributions including their total uncertainties.}
\footnotesize
% The next lines are included from the .//Tables/tab_Data.tex input file
\begin{tabular}{l | c | c}
\toprule
& \Zgee 	& \Zgmm 	\\
\midrule
Two reconstructed leptons within fiducial volume	&13\,649\,239			& 18\,162\,641				\\
\midrule
Electroweak background ($Z\rightarrow\tau\tau$, $WW$, $WZ$, $ZZ$) &   ~$40\,000 \pm 2000$  & ~$50\,000 \pm 2500$	\\
Photon-induced	 background	&   ~~$2900 \pm 140$ 	  & ~~$4100  \pm 200$	\\
Top-quark background		&   ~$38\,000 \pm 1900$  & ~$45\,400 \pm 2200$	\\
Multijet background 	        &   ~~~~$8500 \pm 4900$	   & ~~$1000 \pm 200$	\\
\midrule
& $89\,400\pm5600$ & $100\,500\pm3300$\\
\bottomrule
\end{tabular}
% End of text imported from the .//Tables/tab_Data.tex input file
\centering
\end{table}
 
\subsection{Estimation of backgrounds}
\label{sec:backgrounds}
 
The backgrounds from all sources other than multijet processes are estimated using the MC samples detailed in Section~\ref{sec:datamc}.
The number and properties of the background events where one or two reconstructed lepton candidates originate from hadrons or hadron decay products,
i.e.\ multijet processes as well as $W$+jets, are estimated using the data-driven techniques described in the following for both decay channels.
 
In the electron channel, a multijet-dominated sample is selected from
data with two same-charge electron candidates satisfying the `loose'
identification criteria, but not the `medium'
criteria~\cite{PERF-2017-01}, i.e.\ they are more likely to be caused by misidentified jets.
This sample is collected by a di-electron trigger without isolation criteria~\cite{TRIG-2018-05}.
In the muon channel, a multijet sample
is obtained by selecting two same-charge muons. The residual
contamination from processes with prompt leptons is estimated using
the simulation and subtracted.
 
The normalization of the multijet template in the electron channel is determined in a fit to the distribution of the electron isolation using all event-selection criteria except those for the isolation variables. Systematic uncertainties in the normalization are estimated by varying the fit range on the electron isolation distribution.
 
In the muon channel, the normalization is obtained using the ratio of
number of opposite-charge dimuon events to the number of same-charge dimuon events where the muons
fail to satisfy the isolation criterion. Assuming no correlation between the
isolation of muons in multijet events and their charge, this ratio can be
applied to a control sample, defined by pairs of isolated same-charge muons passing the signal-kinematic selection, to determine the multijet contamination in the signal region. The systematic uncertainty in the estimate is
obtained by varying the isolation criterion for the muons.
 
The total fraction of selected data events originating from background
processes is about 0.6\% in both the electron and muon
channels. The background is dominated by contributions from diboson
and \ttbar processes. An overview of the estimated number of background events is given in Table \ref{tab:DataZleptons}, together with the corresponding total uncertainties.
 
\begin{figure}[htbp]
\begin{center}
\includegraphics[width=0.495\textwidth]{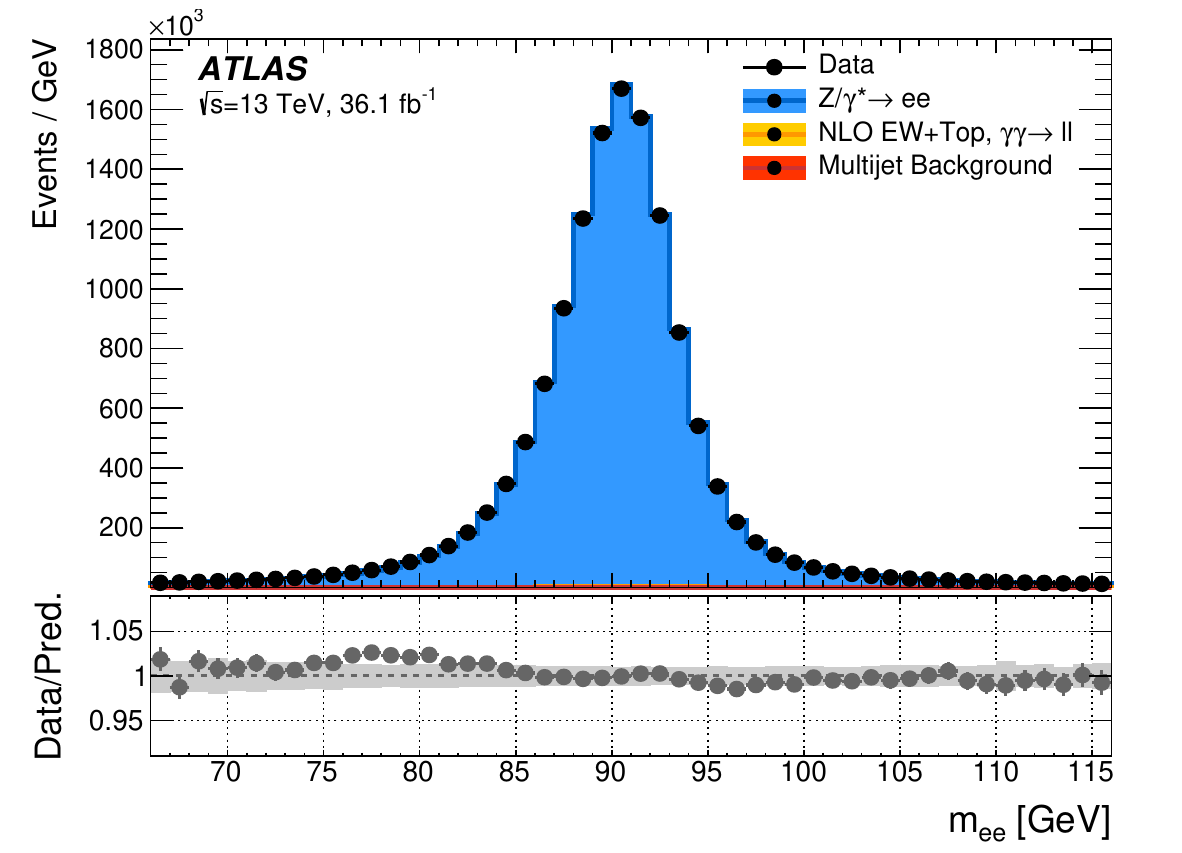}
\includegraphics[width=0.495\textwidth]{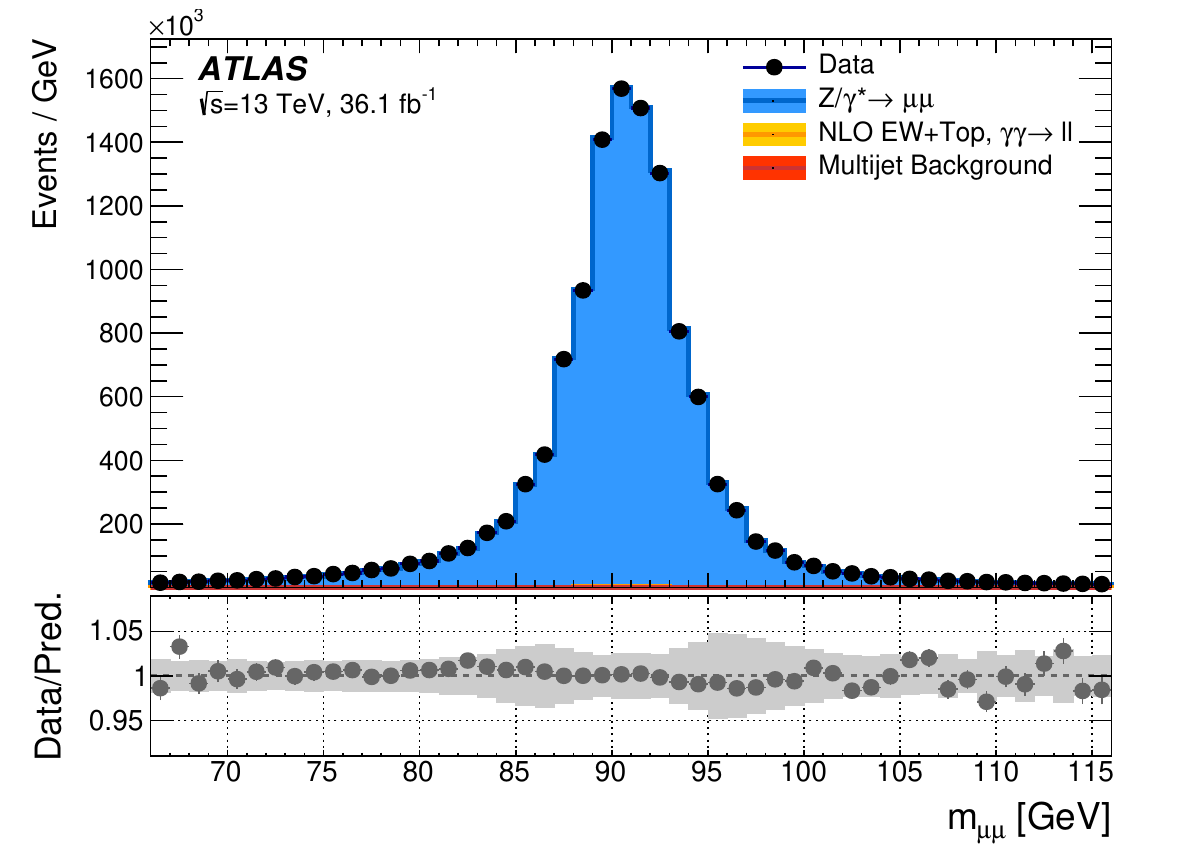}
\includegraphics[width=0.495\textwidth]{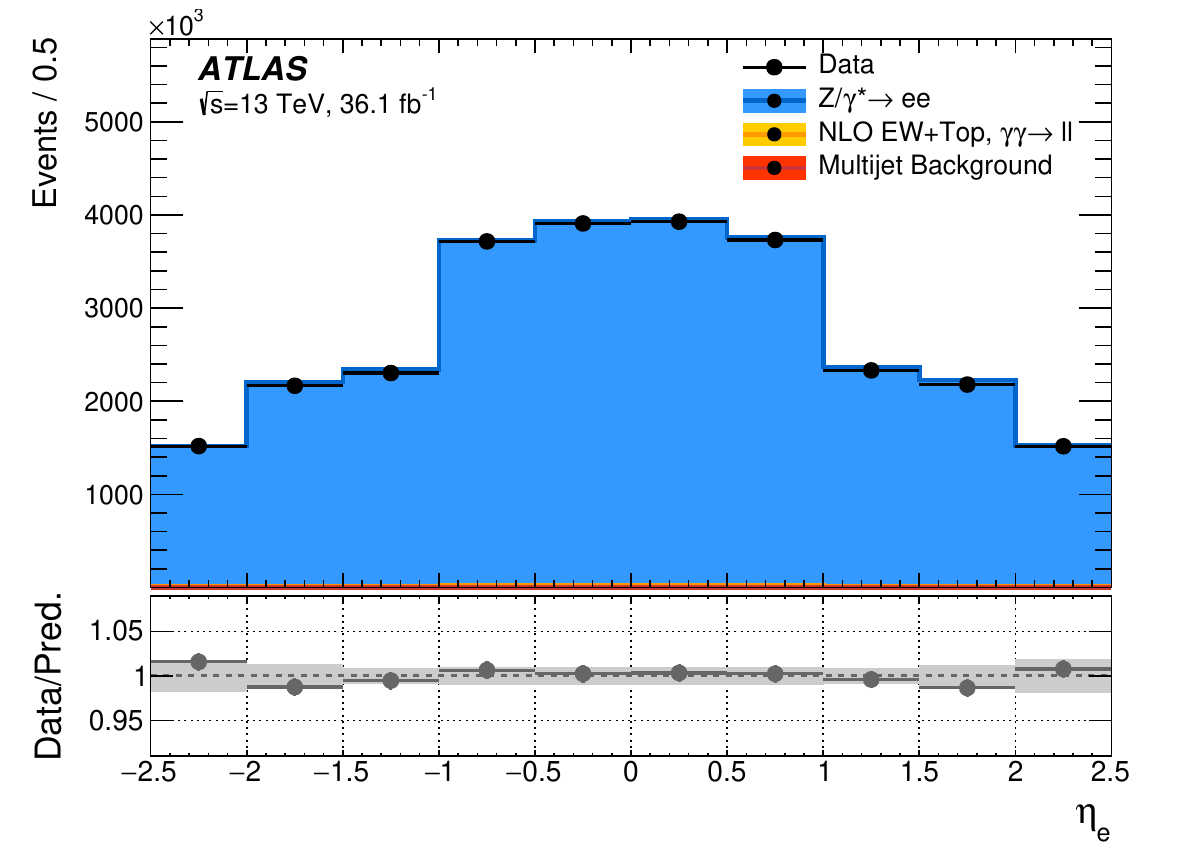}
\includegraphics[width=0.495\textwidth]{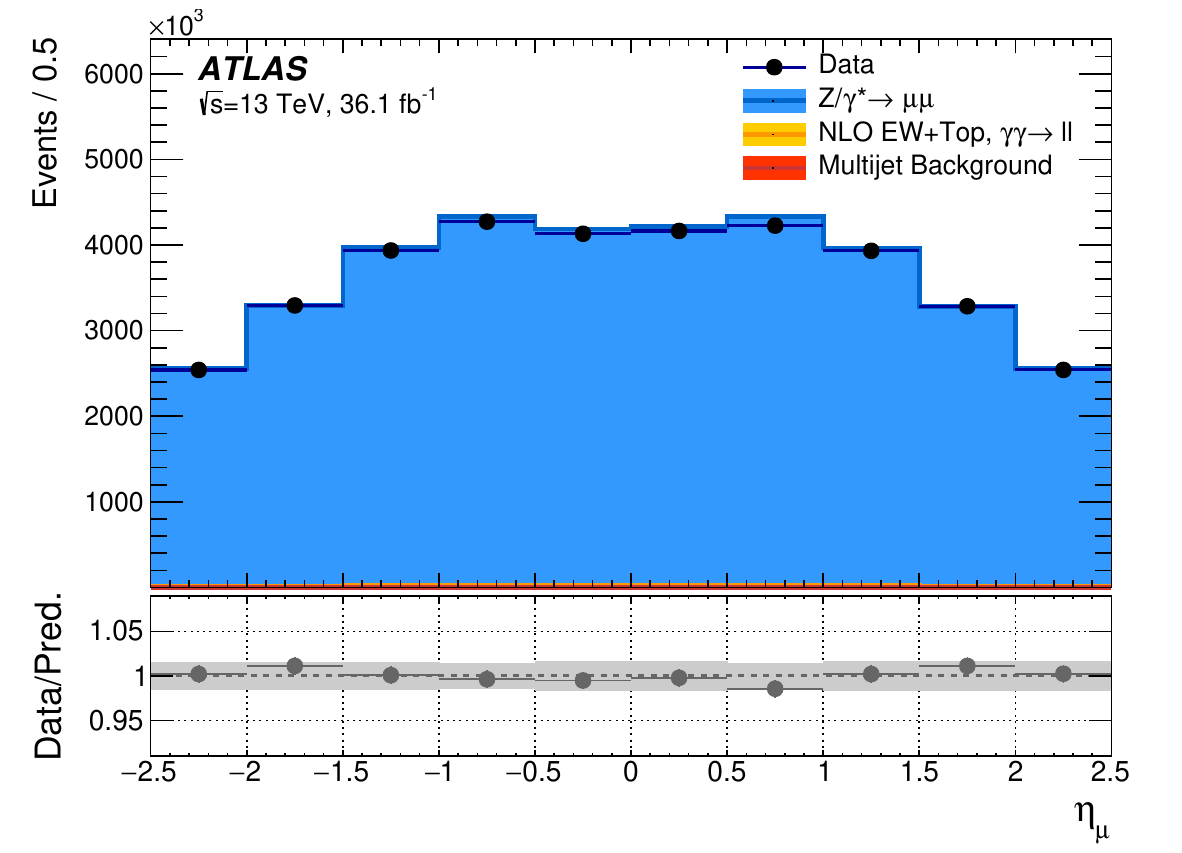}
\caption{
\label{fig:controlplots}
The distribution of events passing the selection requirements in the electron channel (left) and muon channel (right) as a function of dilepton invariant mass \Mll (upper row) and lepton pseudorapidity $\eta$ (lower row), the latter with one entry for each lepton per event. The MC signal sample is simulated using \POWHEG{}+\PYTHIA{}8. The statistical uncertainties of the data points are generally smaller than the size of the markers. The predictions of the MC signal sample together with the MC background samples are normalized to the integral of the data and the total experimental uncertainty of the predicted values is shown as a grey band in the ratio of the prediction to data.}
\end{center}
\end{figure}

\begin{figure}[htbp]
\begin{center}
\includegraphics[width=0.495\textwidth]{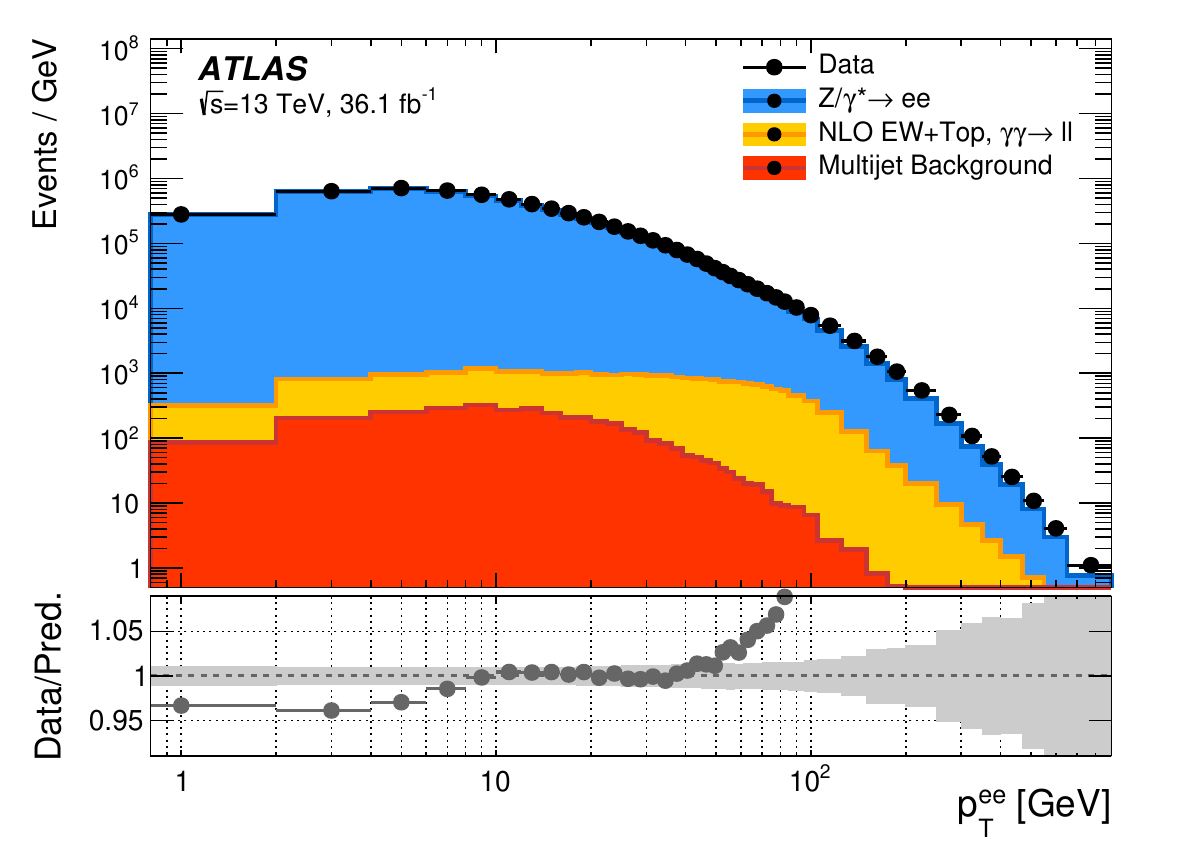}
\includegraphics[width=0.495\textwidth]{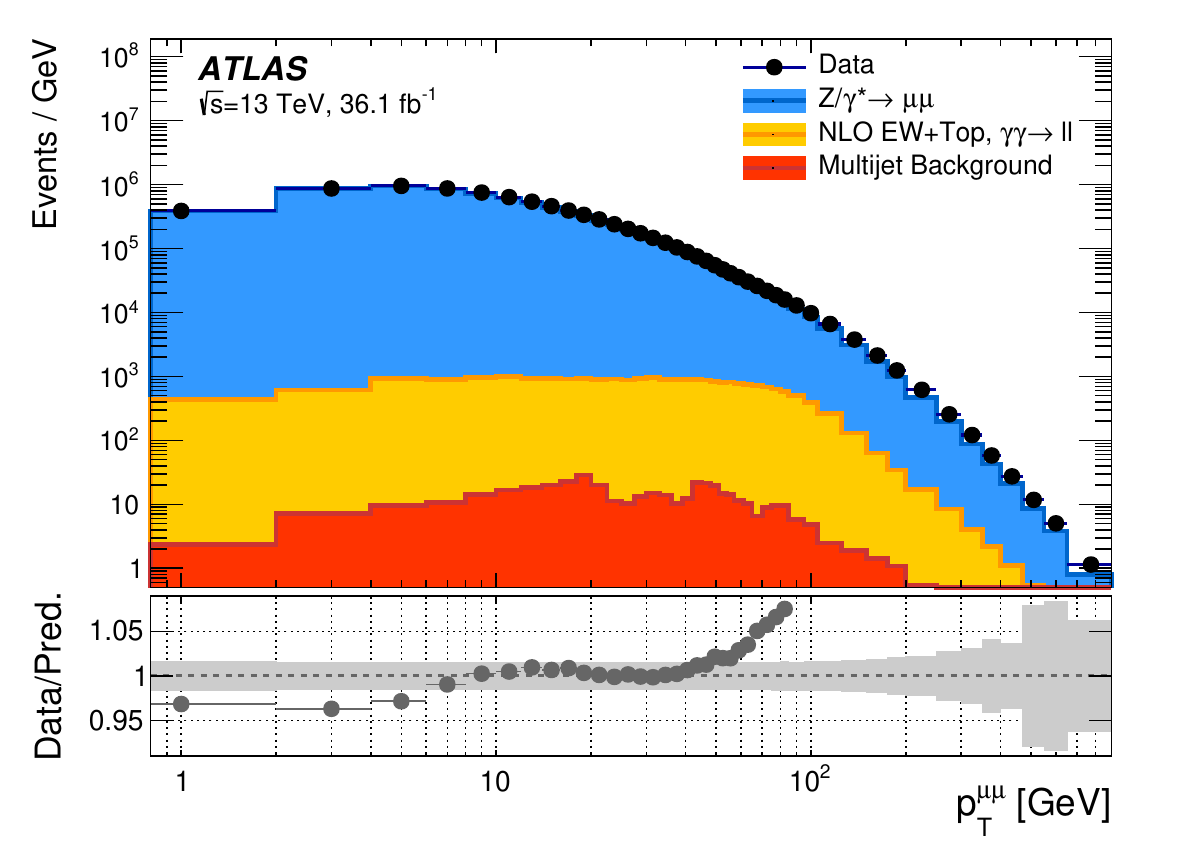}
\includegraphics[width=0.495\textwidth]{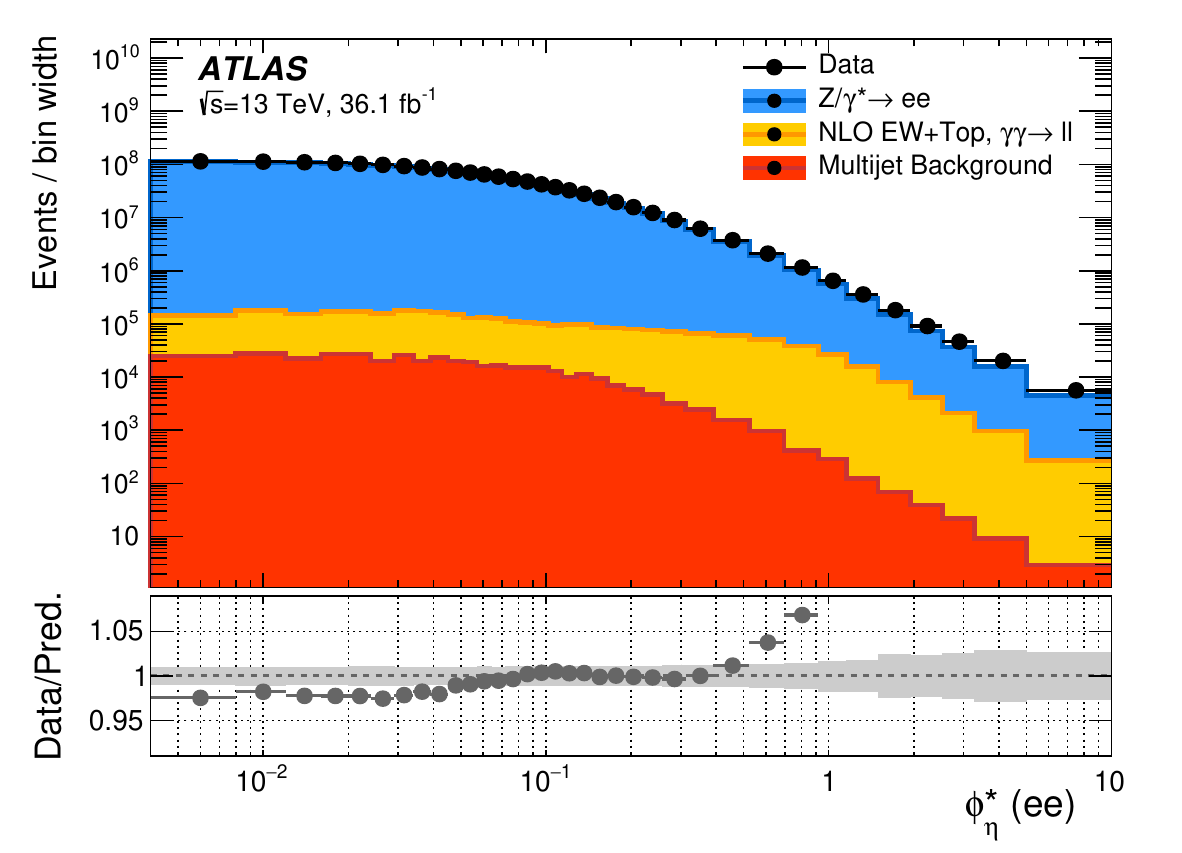}
\includegraphics[width=0.495\textwidth]{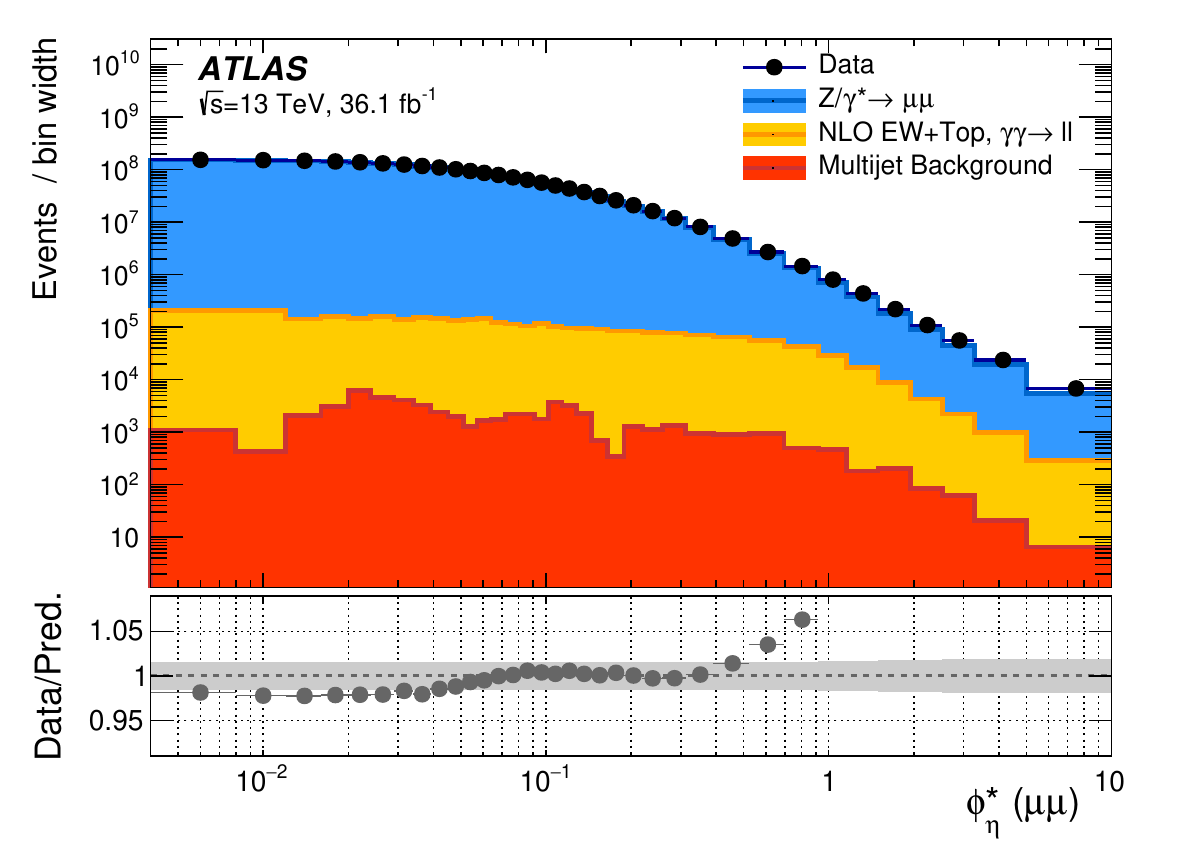}
\caption{
\label{fig:controlplotsUnfolding}
The distribution of events passing the selection requirements in the electron channel (left) and muon channel (right) as a function of dilepton transverse momentum (upper row) and \phiStar{} (lower row). The MC signal sample is simulated using \POWHEG{}+\PYTHIA{}8. The statistical uncertainties of the data points are generally smaller than the size of the markers. The predictions are normalized to the integral of the data and the total experimental uncertainty of the predicted values is shown as a grey band in the ratio of the prediction to data.}
\end{center}
\end{figure}

Figure~\ref{fig:controlplots} shows the dilepton invariant mass and the lepton pseudorapidity distribution, for the electron and muon channels separately. The predictions are in fair agreement with the data. The impact of the residual differences between these distributions on the \Zpt{} and \phiStar{} measurements is estimated by reweighting the MC signal sample to data and then repeating the measurement procedure. Figure~\ref{fig:controlplotsUnfolding} compares the measured  \Zpt{} and \phiStar{} distributions for both channels with the signal MC predictions. The disagreement between the data and the predictions for large values of \Zpt{} and \phiStar{} is expected because \POWHEG{}+\PYTHIA{}8 is effectively a computation at leading-order in $\alphas$ in this region.
 
\subsection{Correction for detector effects}
 
The production cross-section times the branching ratio for decays into
a single lepton flavour are measured in fiducial volumes as defined in
Section \ref{sec:desc}. Integrated fiducial cross-sections in the electron and muon channels
are computed following the equation
 
\begin{equation*}
\sigma^\mathrm{fid}_{\Zgll} = \frac{N_\mathrm{Data}-N_\mathrm{Bkg}}{C_Z \cdot L},
\end{equation*}
 
where $N_\mathrm{Data}$ is the number of observed signal candidates and $N_\mathrm{Bkg}$ is the number of background events expected in the selected sample. The integrated luminosity of the sample is $L = 36.1$\,\ifb. A correction for the event detection efficiency is applied with the factor $C_Z$, which is obtained from the simulation of signal events as the ratio of the sum of event weights after simulation, reconstruction and selection, to the sum of MC event weights for events satisfying the fiducial requirements. The factor $C_Z$ is affected by experimental uncertainties, described in Section \ref{sec:Systematics}, while theory and modelling uncertainties are negligible.
 
The differential distributions within the fiducial volume are corrected for detector effects and bin-to-bin migrations using an iterative Bayesian unfolding method~\cite{DAgostini:1994zf,DAgostini:2010,Adye:2011gm}. First, the data are corrected for events that pass the detector-level selection but not the particle-level selection. Then, the iterative Bayesian unfolding technique is used as a regularized way to correct for the detector resolution in events that pass both the detector-level and particle-level selections. The method is applied with four iterations implemented in the RooUnfold framework \cite{Adye:2011gm}. After the application of the response matrix, a final correction is applied to account for events that pass the particle-level but not detector-level selection, resulting in unfolded distributions on Born and dressed particle level. The response matrices, which connect the distributions at reconstruction and particle level, as well as the correction factors are derived using the \POWHEG{}+\PYTHIA signal MC sample.
% End of text imported from the .//Sec_3_AnalysisMethodology.tex input file
% The next lines are included from the .//Sec_4_Systematics.tex input file
\section{\label{sec:Systematics}Statistical and systematic uncertainties}
 
Uncertainties in the measurement are assessed for each aspect of the analysis, including the background subtraction, event detection efficiencies, response matrix, and unfolding method.  The entire analysis procedure is repeated for each systematic uncertainty. For each uncertainty, the corresponding source of uncertainty is varied to estimate the effect on the final result.
 
The effect on the measurement from the size of the data and MC samples is estimated by generating pseudo-experiment variations of the respective samples. The resulting statistical uncertainties are considered as uncorrelated between bins and between channels.
 
Uncertainties in the scale and resolution of the electron energy scale~\cite{PERF-2017-03} and muon momentum scale~\cite{PERF-2015-10} are among the dominant uncertainties in the \Zpt{} measurement. Furthermore, uncertainties related to lepton reconstruction and selection efficiencies are considered~\cite{PERF-2017-01,PERF-2015-10,TRIG-2016-01, TRIG-2018-05}, covering the lepton identification, reconstruction, isolation, triggering and track-to-vertex matching processes. The lepton related systematic uncertainties have only a small statistical component. There is an additional uncertainty in the muon channel to cover charge-dependent biases in the muon momentum measurement. The majority of these experimental uncertainties are considered correlated between bins of \Zpt{} and \phiStar{}. An exception are the components of the reconstruction and identification efficiencies which have a significant statistical component due to the limited number of events in the data samples used to derive the efficiency corrections. Uncertainties related to electron or muon reconstruction and identification are always assumed to be uncorrelated with each other. They dominate the uncertainty in the fiducial cross-section measurement.
 
The uncertainties in the MC background estimates are obtained by independently varying the theory cross-sections used to normalize the corresponding samples and observing the effect on the measured \Zpt{} and \phiStar{} cross-sections. These background uncertainties are considered correlated between bins of \Zpt{} and \phiStar{} and between the electron and muon channels. As described in Section~\ref{sec:backgrounds}, the uncertainty in the multijet background in the electron channel is obtained by changing the input range of the template used to estimate the multijet background. For the muon channel, the measurement is performed again with a modified isolation variable used in the normalization procedure. The differences between the nominal and modified measurements are used as uncertainty. The estimated multijet backgrounds are assumed to be uncorrelated between the channels.
 
An uncertainty is derived to cover the mis-modelling of the simulated pile-up activity following the measurement of the cross-section of inelastic $pp$ collisions~\cite{STDM-2015-05}. Also, the uncertainty in modelling the distribution of the longitudinal position of the primary vertex is considered. These uncertainties are treated as correlated between the electron channel and muon channel.
 
The uncertainty from the unfolding method is determined by repeating the procedure with a different simulation where the nominal particle-level spectrum is reweighted so that the simulated detector-level spectrum is in good agreement with the data.
The modified detector-level distribution is unfolded with the nominal response matrix and the difference between the result and the reweighted particle-level spectrum is taken as the bias of the unfolding method due to the choice of prior. 
The closure of the unfolding procedure is also tested using the generator-level distributions of the \SHERPA{} MC sample described in Section~\ref{sec:comparepred}, where consistent results within the assigned unfolding uncertainties are found.
 
The uncertainty from the choice of PDF used in the signal MC generator is evaluated by reweighting the signal MC simulation to the 52 error sets of the CT10 PDF set and computing the resulting variation of the results~\cite{Lai:2010vv,Buckley:2014ana}. The differences found in this way are negligible, similar to scale-choice uncertainties. The uncertainty in the combined 2015--2016 integrated luminosity is 2.1\% \cite{ATLAS-CONF-2019-021}, obtained using the LUCID-2 detector \cite{LUCID2} for the primary luminosity measurements. This uncertainty only applies to the absolute cross-section measurements.

\begin{table}[tb]
\caption{\label{tab:CFacUncertainty}Overview of the detector efficiency correction factors, $C_Z$, for the electron and muon channels and their systematic uncertainty contributions.}
\footnotesize
\centering
% The next lines are included from the .//Tables/tab_Uncertainties.tex input file
 
\begin{tabular}{l | cc | cc}
\hline
& \multicolumn{2}{c}{Electron channel} 	& \multicolumn{2}{c}{Muon channel}	\\
\hline
& Born	& Dressed					& Born 	& Dressed			\\
\hline
$C_Z$					& $0.509\pm 0.005$	& $0.522\pm 0.005$	& $0.685\pm 0.011$	& $0.702\pm 0.011$\\
\hline
Trigger efficiencies						& \multicolumn{2}{c|}{$\pm$\,0.0004}& \multicolumn{2}{c}{$\pm$\,0.0004}	\\
Identification \& reconstruction efficiencies		& \multicolumn{2}{c|}{$\pm$\,0.0049}	& \multicolumn{2}{c}{$\pm$\,0.0102}	\\
Isolation efficiencies						& \multicolumn{2}{c|}{$\pm$\,0.0009}	& \multicolumn{2}{c}{$\pm$\,0.0029}	\\
Energy/momentum scale and resolution		& \multicolumn{2}{c|}{$\pm$\,0.0014}	& \multicolumn{2}{c}{$\pm$\,0.0010}	\\
Pile-up								& \multicolumn{2}{c|}{$\pm$\,0.0011}	& \multicolumn{2}{c}{$\pm$\,0.0019}	\\
Model uncertainties						& \multicolumn{2}{c|}{$\pm$\,0.0001}	& \multicolumn{2}{c}{$\pm$\,0.0001}	\\
\hline
\end{tabular}

% End of text imported from the .//Tables/tab_Uncertainties.tex input file
\end{table}
 
The experimental uncertainties of $C_Z$ for the integrated fiducial cross-section
measurements in the electron and muon channels are summarized
in Table~\ref{tab:CFacUncertainty}. The electron identification
efficiency and muon reconstruction efficiency contribute a large
fraction of the total systematic uncertainty for both the integrated
and absolute differential measurements. These uncertainties are
greatly reduced for the normalized measurement of differential
distributions. A summary of the uncertainties in the normalized differential
cross-section measurements is provided in Figure~\ref{fig:systematics}
as a function of \Zpt{} and \phiStar{} for both decay channels.
The statistical uncertainties for the electron and muon channel measurements are a combination of the uncertainties due to limited data and MC sample sizes.
The systematic uncertainties are divided into categories and originate from lepton scales and resolutions, reconstruction and identification efficiencies, as well as the MC signal modelling in the unfolding procedure and further smaller uncertainty sources such as the subtraction of background contributions. These smaller contributions are summarized as "other" uncertainties.

\begin{figure}[htbp]
\begin{center}
\includegraphics[width=0.495\textwidth]{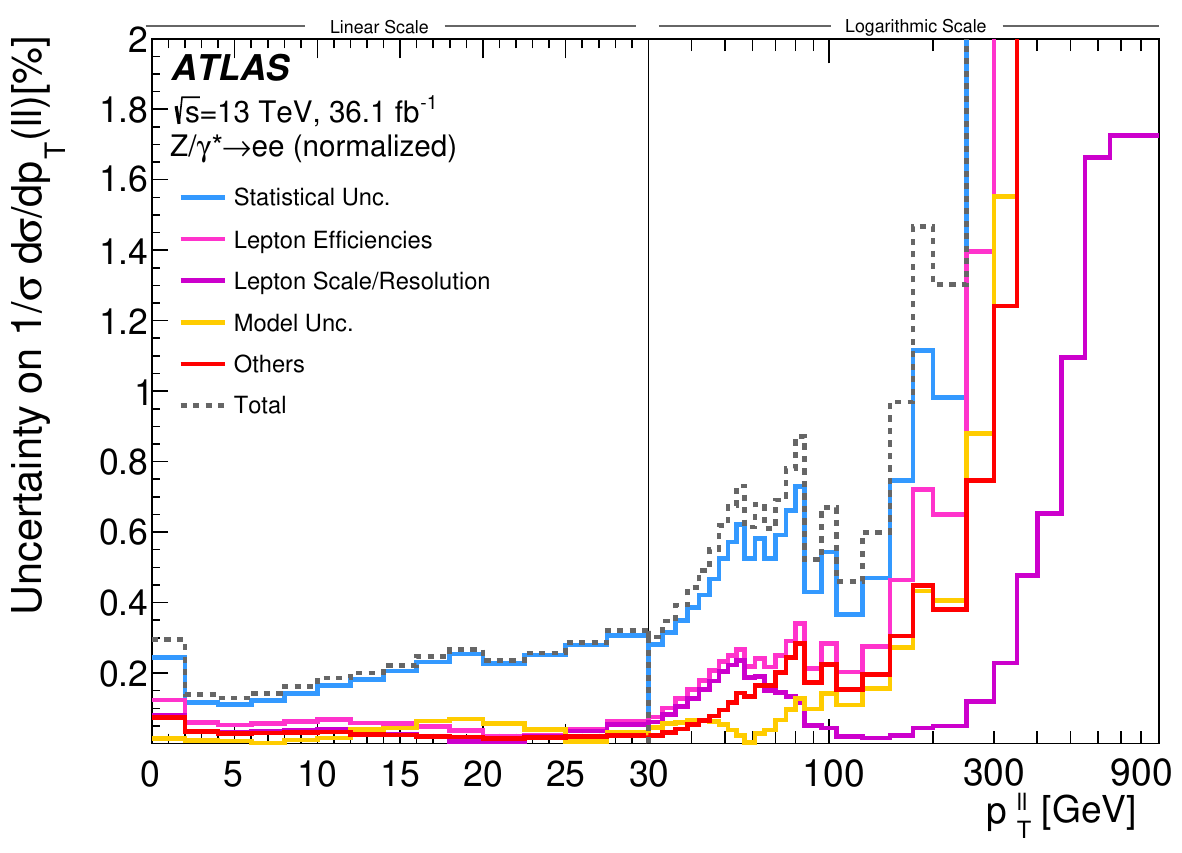}
\includegraphics[width=0.495\textwidth]{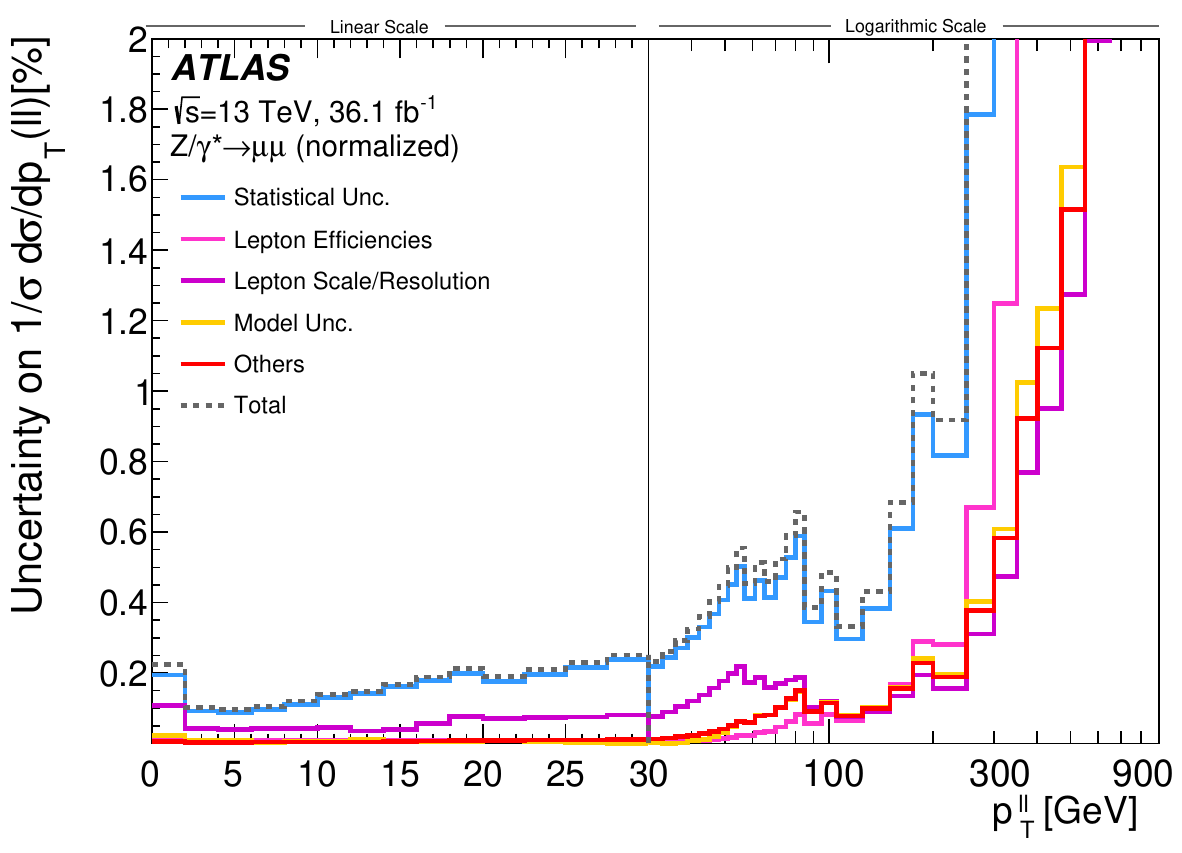}
\includegraphics[width=0.495\textwidth]{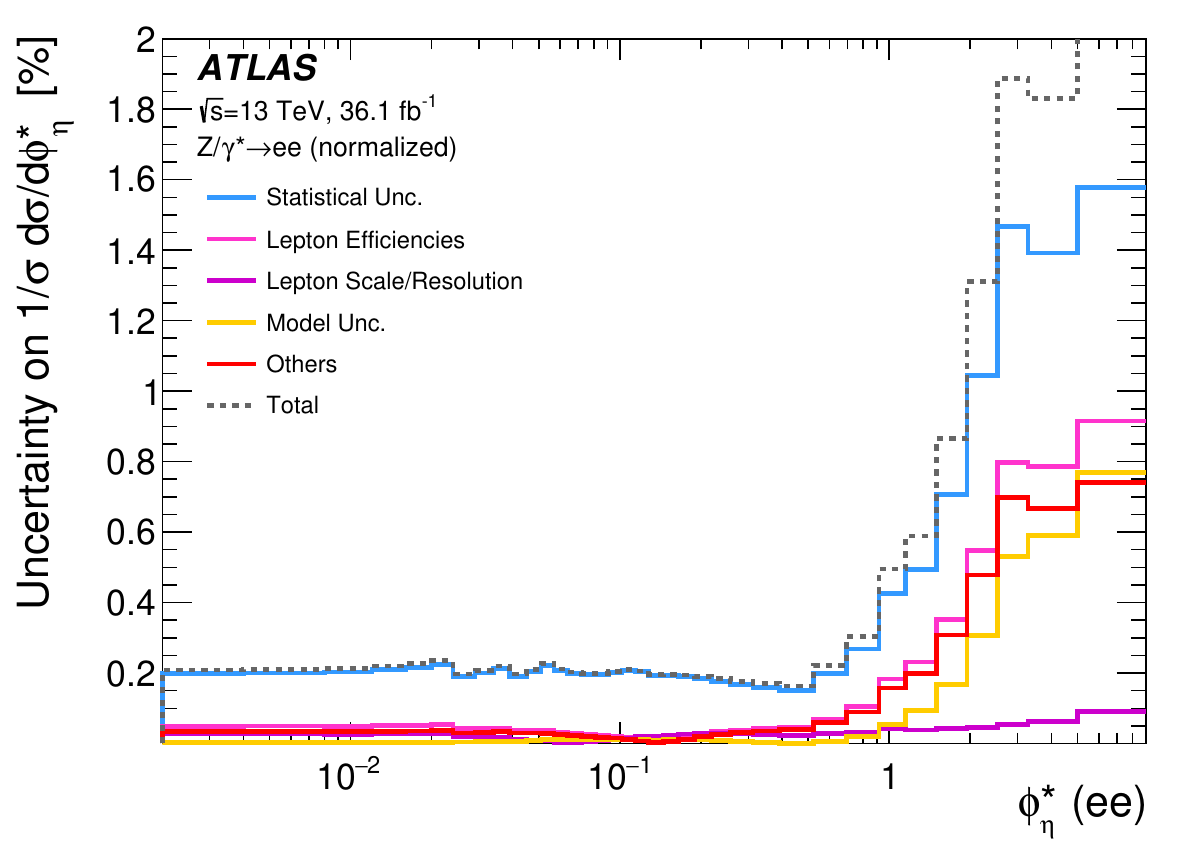}
\includegraphics[width=0.495\textwidth]{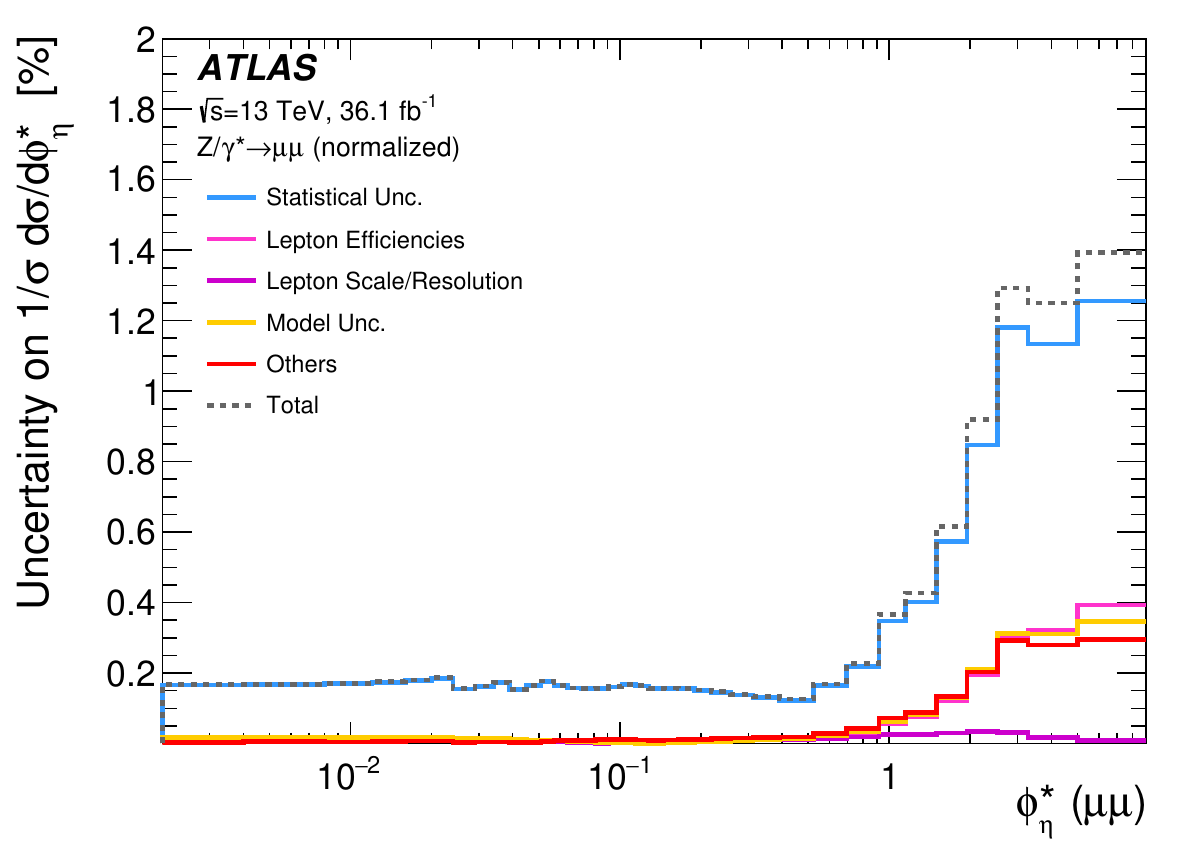}
\caption{
\label{fig:systematics}
The systematic uncertainties for the electron channel measurement (left) and muon channel measurement (right) for the normalized \Zpt{} (upper row) and normalized \phiStar{} (lower row). The statistical uncertainties are a combination of the uncertainties due to limited data and MC sample sizes. The $\Zpt{}$ distribution is split into linear and logarithmic scales at 30~\GeV. Some uncertainties are larger than 2\% for $\Zpt{}>200$ GeV and hence cannot be displayed. The corresponding uncertainties are also summarized in Table \ref{tab:difcrosssections}.
}
\end{center}
\end{figure}
% End of text imported from the .//Sec_4_Systematics.tex input file
% The next lines are included from the .//Sec_5_Results.tex input file
\section{Results and discussion} 
 
\subsection{Combination}
 
The fiducial cross-sections measured in the \Zgee and \Zgmm channels are presented in Table \ref{tab:crosssections} including statistical, systematic and luminosity uncertainties. When correcting for the more restrictive fiducial volume definition, these results are in good agreement with the previous ATLAS measurements at 13~\TeV~\cite{STDM-2016-02}. The electron- and muon-channel cross-sections are combined using $\chi^2$ minimization, following the best linear unbiased estimator prescription (\textsc{Blue}) \cite{Lyons:1988rp, Valassi:2003mu, Nisius:2014wua}. The combination is performed on Born level, resulting in a combined cross-section of $\sigma_\mathrm{fid} (pp\rightarrow \Zgll) =$ 736.2 $\pm$ 0.2(stat) $\pm$ 6.4 (sys) $\pm$ 15.5 (lumi)~pb (Table~\ref{tab:crosssections})\footnote{The results on dressed level are about 2.4\% lower compared to the Born level definition}. There is a reduction of the uncertainty compared to individual electron- and muon-channel measurements since the dominant detector-related systematic uncertainty sources are largely uncorrelated. The uncertainties due to pile-up, physics modelling and luminosity are treated as correlated between the two decay channels.
 
\begin{table}[thb]
\caption{\label{tab:crosssections}Measured integrated cross-section in the fiducial volume in the electron and muon decay channels at Born level and their combination as well as the theory prediction at NNLO in $\alphas$ using the CT14 PDF set.}
\footnotesize
\centering
% The next lines are included from the .//Tables/tab_CrossSections.tex input file
\begin{tabular}{l | c | c }
\hline
Channel					&	Measured cross-section $\times$ $\mathcal{B}$(\Zgll) 		&	Predicted cross-section $\times$ $\mathcal{B}$(\Zgll) 		\\
&	(value $\pm$ stat. $\pm$ syst. $\pm$ lumi.) 					&	(value $\pm$ PDF $\pm$ $\alphas$ $\pm$ scale $\pm$ intrinsic)				\\
\hline
\Zgee		&	$738.3 \pm 0.2  \pm \hspace{0.18cm}7.7  \pm 15.5 $\,pb		&													\\
\Zgmm		&	$731.7 \pm 0.2  \pm 11.3  \pm 15.3 $\,pb	&													\\
\Zgll			& 	$736.2 \pm 0.2  \pm \hspace{0.18cm}6.4  \pm 15.5 $\,pb	&	$703^{+19}_{-24} \; ^{+6}_{-8} \; ^{+4}_{-6} \; ^{+5}_{-5}$\,pb	\cite{STDM-2016-02}	\\
\hline
\hline
\end{tabular}
% End of text imported from the .//Tables/tab_CrossSections.tex input file
\end{table}
 
The normalized differential cross-sections $1/\sigma_\mathrm{fid} \times \mathrm{d}\sigma_\mathrm{fid}/\mathrm{d}\Zpt{}$ and $1/\sigma_\mathrm{fid} \times \mathrm{d}\sigma_\mathrm{fid}/\mathrm{d}\phiStar{}$ measured in the two decay channels as well as their combination are illustrated in Figure~\ref{fig:combined}. When building the $\chi^2$ for combination procedure, the measurement uncertainties are separated into those from bin-to-bin uncorrelated sources and those from bin-to-bin correlated sources, the latter largely reduced due to the normalization by the fiducial cross-section. The normalized differential measurements are combined at Born level following the \textsc{Blue} prescription. The resulting $\chi^2/N_\mathrm{dof} = 47/44$ for the combination for \Zpt{} and the $\chi^2/N_\mathrm{dof} = 32/36$ for \phiStar{} indicate good agreement between the two channels.\footnote{The $\chi^2/N_\mathrm{dof}$ is still good when taking into account only bins with $\Zpt{}>50$~\GeV.} The combined precision is between 0.1\% and 0.5\% for $\Zpt{} < 100$~\GeV, rising to 10\% towards the high end of the spectrum, where the overall precision is limited by the data and MC sample size. The combined results for both distributions are presented in Table~\ref{tab:difcrosssections} including statistical and bin-to-bin uncorrelated and correlated systematic uncertainties. The measurement results are reported at Born level and factors $k_\mathrm{dr}$, the binwise ratio of dressed and born level results, are given to transfer to the dressed particle level.
 
\begin{table}[h!]
\caption{\label{tab:difcrosssections}The measured combined normalized differential cross-sections, divided by the bin-width, in the fiducial volume at Born level as well as a factor $k_\mathrm{dr}$ to translate from the Born particle level to the dressed particle level.}
\footnotesize
\centering
% The next lines are included from the .//Tables/tab_DiffCrossSections_DivBinWidth.tex input file
\begin{tabular}{lllll | llllll}
\hline
Bin			& $1/\sigma_\mathrm{fid}\,\times$ 		& Corr. 	& Uncorr. 	& $k_\mathrm{dr}$	& Bin	& $1/\sigma_\mathrm{fid}\,\times$ 	& Corr. 	& Uncorr. 	& $k_\mathrm{dr}$	\\
$[$\GeV$]$	& $\mathrm{d}\sigma/\mathrm{d}\Zpt{}$	& uncert. 		& uncert.		&			& 	& $\mathrm{d}\sigma/\mathrm{d}\phiStar{}$ & uncert. 		& uncert.		&			\\
& $[$1/GeV$]$					&			&			&			&		&						&			&			& 	\\
\hline
0-2		& 0.024189	& $\pm$ 0.15\%	&$\pm$ 0.18\%	& 0.978	& 0-0.004		& 8.8053	& $\pm$ 0.03\%	&$\pm$ 0.13\% 	&0.992	\\
2-4		& 0.051144	& $\pm$ 0.06\%	&$\pm$ 0.08\%	& 0.985	& 0.004-0.008	& 8.6969	& $\pm$ 0.03\%	&$\pm$ 0.13\% 	&0.993	\\
4-6		& 0.053232	& $\pm$ 0.05\%	&$\pm$ 0.08\%	& 0.994	& 0.008-0.012	& 8.5624	& $\pm$ 0.02\%	&$\pm$ 0.13\% 	&0.993	\\
6-8		& 0.047383	& $\pm$ 0.05\%	&$\pm$ 0.08\%	& 1.000	& 0.012-0.016	& 8.3378	& $\pm$ 0.02\%	&$\pm$ 0.13\% 	&0.994	\\
8-10		& 0.040568	& $\pm$ 0.04\%	&$\pm$ 0.09\%	& 1.010	& 0.016-0.02	& 8.0881	& $\pm$ 0.03\%	&$\pm$ 0.14\% 	&0.994	\\
10-12	& 0.034317	& $\pm$ 0.06\%	&$\pm$ 0.11\%	& 1.010	& 0.02-0.024	& 7.7920	& $\pm$ 0.03\%	&$\pm$ 0.14\% 	&0.995	\\
12-14	& 0.029157	& $\pm$ 0.07\%	&$\pm$ 0.12\%	& 1.010	& 0.024-0.029	& 7.4174	& $\pm$ 0.02\%	&$\pm$ 0.12\% 	&0.995	\\
14-16	& 0.024804	& $\pm$ 0.06\%	&$\pm$ 0.14\%	& 1.010	& 0.029-0.034	& 7.0360	& $\pm$ 0.02\%	&$\pm$ 0.13\% 	&0.996	\\
16-18	& 0.021268	& $\pm$ 0.05\%	&$\pm$ 0.15\%	& 1.010	& 0.034-0.039	& 6.5989	& $\pm$ 0.02\%	&$\pm$ 0.13\% 	&0.998	\\
18-20	& 0.018325	& $\pm$ 0.04\%	&$\pm$ 0.16\%	& 1.010	& 0.039-0.045	& 6.1608	& $\pm$ 0.02\%	&$\pm$ 0.12\% 	&0.998	\\
20-22.5	& 0.015605	& $\pm$ 0.03\%	&$\pm$ 0.14\%	& 1.010	& 0.045-0.051	& 5.7085	& $\pm$ 0.01\%	&$\pm$ 0.13\% 	&0.999	\\
22.5-25	& 0.013180	& $\pm$ 0.03\%	&$\pm$ 0.15\%	& 1.000	& 0.051-0.057	& 5.2791	& $\pm$ 0.02\%	&$\pm$ 0.14\% 	&1.000	\\
25-27.5	& 0.011207	& $\pm$ 0.04\%	&$\pm$ 0.17\%	& 1.000	& 0.057-0.064	& 4.8488	& $\pm$ 0.02\%	&$\pm$ 0.13\% 	&1.000	\\
27.5-30	& 0.0095568	& $\pm$ 0.05\%	&$\pm$ 0.19\%	& 0.999	& 0.064-0.072	& 4.4139	& $\pm$ 0.01\%	&$\pm$ 0.12\% 	&1.000	\\
30-33	& 0.0081029	& $\pm$ 0.06\%	&$\pm$ 0.17\%	& 0.998	& 0.072-0.081	& 3.9705	& $\pm$ 0.01\%	&$\pm$ 0.12\% 	&1.000	\\
33-36	& 0.0067881	& $\pm$ 0.08\%	&$\pm$ 0.19\%	& 0.996	& 0.081-0.091	& 3.5515	& $\pm$ 0.01\%	&$\pm$ 0.12\% 	&1.000	\\
36-39	& 0.0057563	& $\pm$ 0.09\%	&$\pm$ 0.21\%	& 0.994	& 0.091-0.102	& 3.1421	& $\pm$ 0.02\%	&$\pm$ 0.13\% 	&1.000	\\
39-42	& 0.0048769	& $\pm$ 0.12\%	&$\pm$ 0.23\%	& 0.993	& 0.102-0.114	& 2.7659	& $\pm$ 0.01\%	&$\pm$ 0.13\% 	&1.000	\\
42-45	& 0.0041688	& $\pm$ 0.12\%	&$\pm$ 0.25\%	& 0.992	& 0.114-0.128	& 2.4125	& $\pm$ 0.01\%	&$\pm$ 0.13\% 	&1.000	\\
45-48	& 0.0035213	& $\pm$ 0.14\%	&$\pm$ 0.28\%	& 0.993	& 0.128-0.145	& 2.0648	& $\pm$ 0.01\%	&$\pm$ 0.12\% 	&1.000	\\
48-51	& 0.0029751	& $\pm$ 0.17\%	&$\pm$ 0.31\%	& 0.991	& 0.145-0.165	& 1.7299	& $\pm$ 0.02\%	&$\pm$ 0.13\% 	&1.000	\\
51-54	& 0.0025433	& $\pm$ 0.18\%	&$\pm$ 0.35\%	& 0.992	& 0.165-0.189	& 1.4282	& $\pm$ 0.02\%	&$\pm$ 0.13\% 	&1.000	\\
54-57	& 0.0021832	& $\pm$ 0.20\%	&$\pm$ 0.38\%	& 0.994	& 0.189-0.219	& 1.1469	& $\pm$ 0.02\%	&$\pm$ 0.12\% 	&1.000	\\
57-61	& 0.0018779	& $\pm$ 0.15\%	&$\pm$ 0.31\%	& 0.994	& 0.219-0.258	& 0.8848	& $\pm$ 0.02\%	&$\pm$ 0.12\% 	&1.000	\\
61-65	& 0.0015932	& $\pm$ 0.17\%	&$\pm$ 0.35\%	& 0.994	& 0.258-0.312	& 0.6470	& $\pm$ 0.03\%	&$\pm$ 0.11\% 	&1.000	\\
65-70	& 0.0013519	& $\pm$ 0.16\%	&$\pm$ 0.32\%	& 0.995	& 0.312-0.391	& 0.4387	& $\pm$ 0.03\%	&$\pm$ 0.11\% 	&1.000	\\
70-75	& 0.0011323	& $\pm$ 0.17\%	&$\pm$ 0.37\%	& 0.995	& 0.391-0.524	& 0.2610	& $\pm$ 0.03\%	&$\pm$ 0.10\% 	&1.000	\\
75-80	& 0.0009574	& $\pm$ 0.20\%	&$\pm$ 0.43\%	& 0.995	& 0.524-0.695	& 0.1414	& $\pm$ 0.04\%	&$\pm$ 0.13\% 	&1.000	\\
80-85	& 0.0008150	& $\pm$ 0.22\%	&$\pm$ 0.49\%	& 0.995	& 0.695-0.918	& 0.07462	& $\pm$ 0.07\%	&$\pm$ 0.17\% 	&1.000	\\
85-95	& 0.0006537	& $\pm$ 0.14\%	&$\pm$ 0.29\%	& 0.996	& 0.918-1.153	& 0.04047	& $\pm$ 0.12\%	&$\pm$ 0.27\% 	&1.000	\\
95-105	& 0.0004849	& $\pm$ 0.18\%	&$\pm$ 0.37\%	& 0.995	& 1.153-1.496	& 0.02167	& $\pm$ 0.14\%	&$\pm$ 0.30\% 	&1.000	\\
105-125	& 0.0003291	& $\pm$ 0.12\%	&$\pm$ 0.25\%	& 0.996	& 1.496-1.947	& 0.01084	& $\pm$ 0.18\%	&$\pm$ 0.42\% 	&1.000	\\
125-150	& 0.0001861	& $\pm$ 0.16\%	&$\pm$ 0.32\%	& 0.994	& 1.947-2.522	& 0.005386	& $\pm$ 0.23\%	&$\pm$ 0.59\% 	&1.000	\\
150-175	& 0.0001050	& $\pm$ 0.24\%	&$\pm$ 0.51\%	& 0.993	& 2.522-3.277	& 0.002738	& $\pm$ 0.31\%	&$\pm$ 0.79\% 	&1.000	\\
175-200	& 6.1279$\cdot 10^{-5}$	& $\pm$ 0.30\%	&$\pm$ 0.78\%	& 0.992	& 3.277-5.000	& 0.0011730	& $\pm$ 0.29\%	&$\pm$ 0.72\% 	&1.000	\\
200-250	& 3.0584$\cdot 10^{-5}$	& $\pm$ 0.22\%	&$\pm$ 0.66\%	& 0.995	& 5.000-10.00	& 0.0003372	& $\pm$ 0.30\%	&$\pm$ 0.78\% 	&0.997	\\
250-300	& 1.2211$\cdot 10^{-5}$	& $\pm$ 0.34\%	&$\pm$ 1.4\%	& 0.997	& 	& \\
300-350	& 5.9026$\cdot 10^{-6}$	& $\pm$ 0.56\%	&$\pm$ 2.3\%	& 0.994	& 	& \\
350-400	& 2.7742$\cdot 10^{-6}$	& $\pm$ 0.90\%	&$\pm$ 3.8\%	& 0.991	& 	& \\
400-470	& 1.2513$\cdot 10^{-6}$	& $\pm$ 0.82\%	&$\pm$ 4.9\%	& 0.991	& 	& \\
470-550	& 5.5219$\cdot 10^{-7}$	& $\pm$ 1.2\%	&$\pm$ 7.9\%	& 0.994	& 	& \\
550-650	& 2.0165$\cdot 10^{-7}$	& $\pm$ 1.5\%	&$\pm$ 13\%	& 0.995	& 	& \\
650-900	& 5.1153$\cdot 10^{-8}$	& $\pm$ 1.8\%	&$\pm$ 16\%	& 0.990	& 	& \\
900-2500	& 1.5735$\cdot 10^{-9}$	& $\pm$ 6.3\%	&$\pm$ 60\%	& 0.964	& 	& \\
\hline
\end{tabular}
% End of text imported from the .//Tables/tab_DiffCrossSections_DivBinWidth.tex input file
\end{table}
 
\begin{figure}[htbp]
\begin{center}
\includegraphics[width=0.495\textwidth]{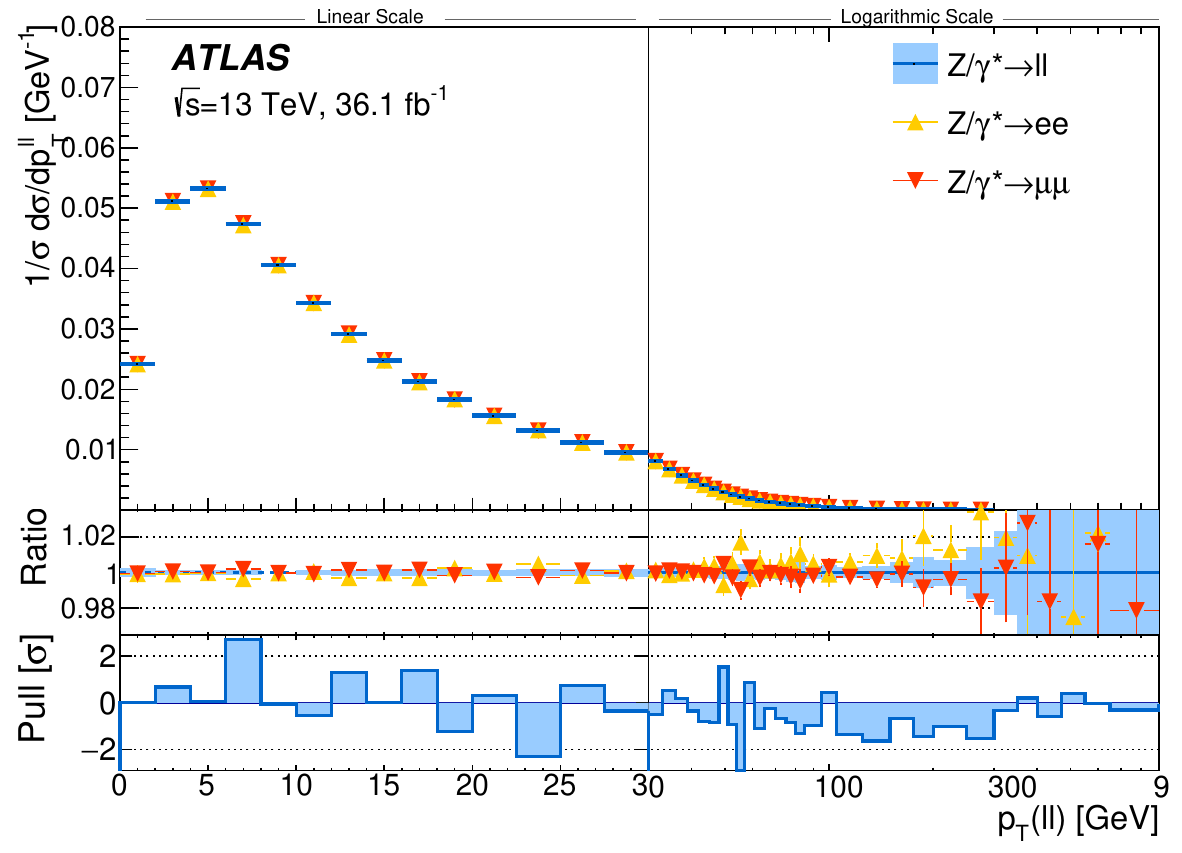}
\includegraphics[width=0.495\textwidth]{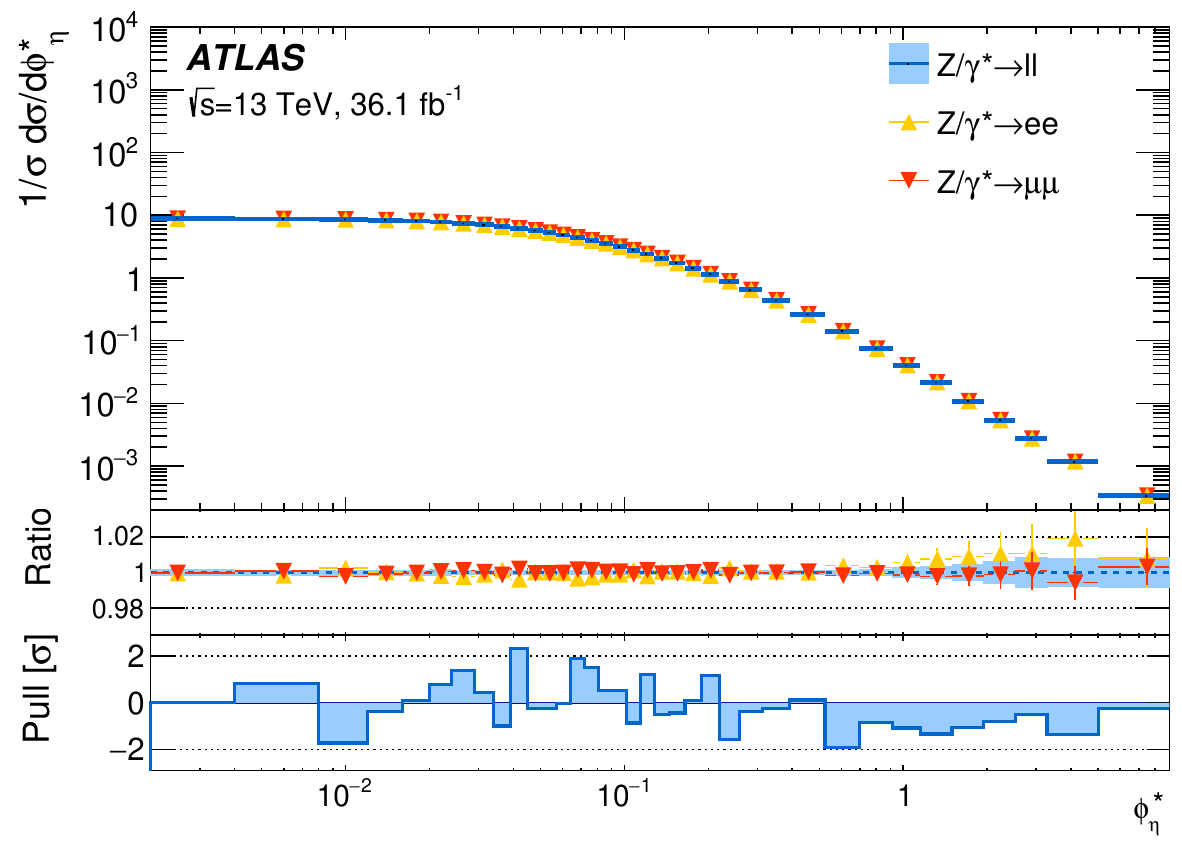}
\caption{\label{fig:combined}The measured normalized cross section as a function of $\Zpt{}$ (left) and \phiStar{} (right) for the electron and muon channels and the combined result as well as their ratio together with the total uncertainties, shown as a blue band. The pull distribution between the electron and muon channels, defined as the difference between the two channels divided by the combined uncorrelated uncertainty, is also shown. The $\Zpt{}$ distribution is split into linear and logarithmic scales at 30~\GeV.}
\end{center}
\end{figure}

\subsection{Comparison with predictions}
\label{sec:comparepred}
 
The integrated fiducial cross-section is compared with a fixed-order
theory prediction that is computed in the same way as in
Ref.~\cite{STDM-2015-03}. The speed-optimized \textsc{DYTurbo}~\cite{Camarda:2019zyx} version of the \textsc{DYNNLO}~1.5~\cite{Catani:2009sm}
program with the CT14 NNLO set of PDFs~\cite{Dulat:2015mca} is used to
obtain a prediction at next-to-next-to-leading order (NNLO) in
$\alphas$ in the $G_\mu$ EW scheme~\cite{Hollik:1988ii}. The
\textsc{FEWZ}~3.1~\cite{Li:2012wna} program is used to compute
next-to-leading-order (NLO) electroweak corrections and to cross-check the
\textsc{DYNNLO} calculation. The prediction is shown in Table
\ref{tab:crosssections} together with its uncertainties estimated as
follows. The dominant uncertainty is from limited knowledge of the proton PDFs
and is estimated using the eigenvectors of the respective CT14 PDF
set, rescaled from 90\% to 68\% confidence level. The uncertainties
due to the strong coupling constant are estimated by varying $\alphas$
by $\pm 0.001$. Missing higher-order QCD corrections are estimated by
variations of the renormalization ($\muR$) and factorization
($\muF$) scales by factors of two with an additional
constraint of $0.5\leq \muR/\muF\leq 2$ around the nominal value of
\Mll. The deviation from the \textsc{FEWZ} calculation is taken as
an intrinsic uncertainty in the NNLO QCD calculation.
A more detailed discussion of the agreement with theory predictions using different PDF sets is given in Ref. \cite{STDM-2016-02}.

\begin{figure}[t]
\begin{center}
\includegraphics[width=0.495\textwidth]{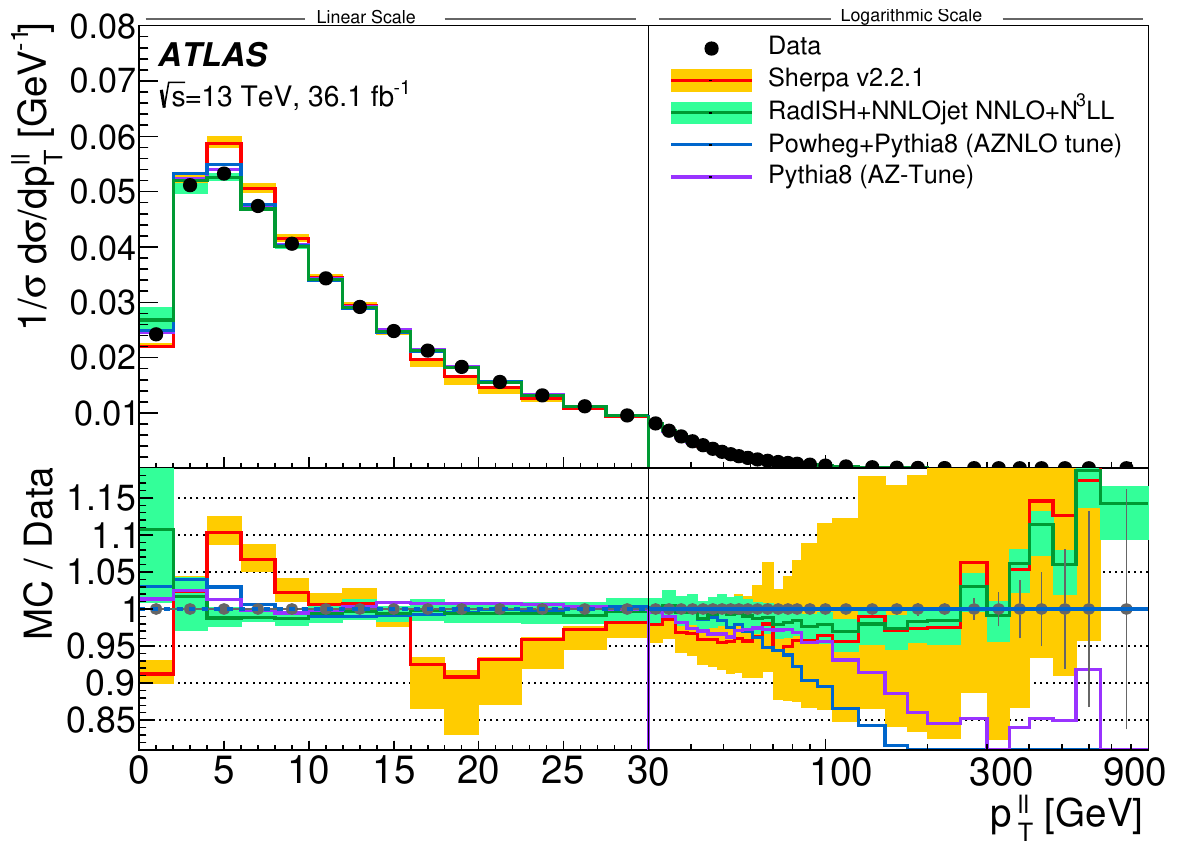}
\includegraphics[width=0.495\textwidth]{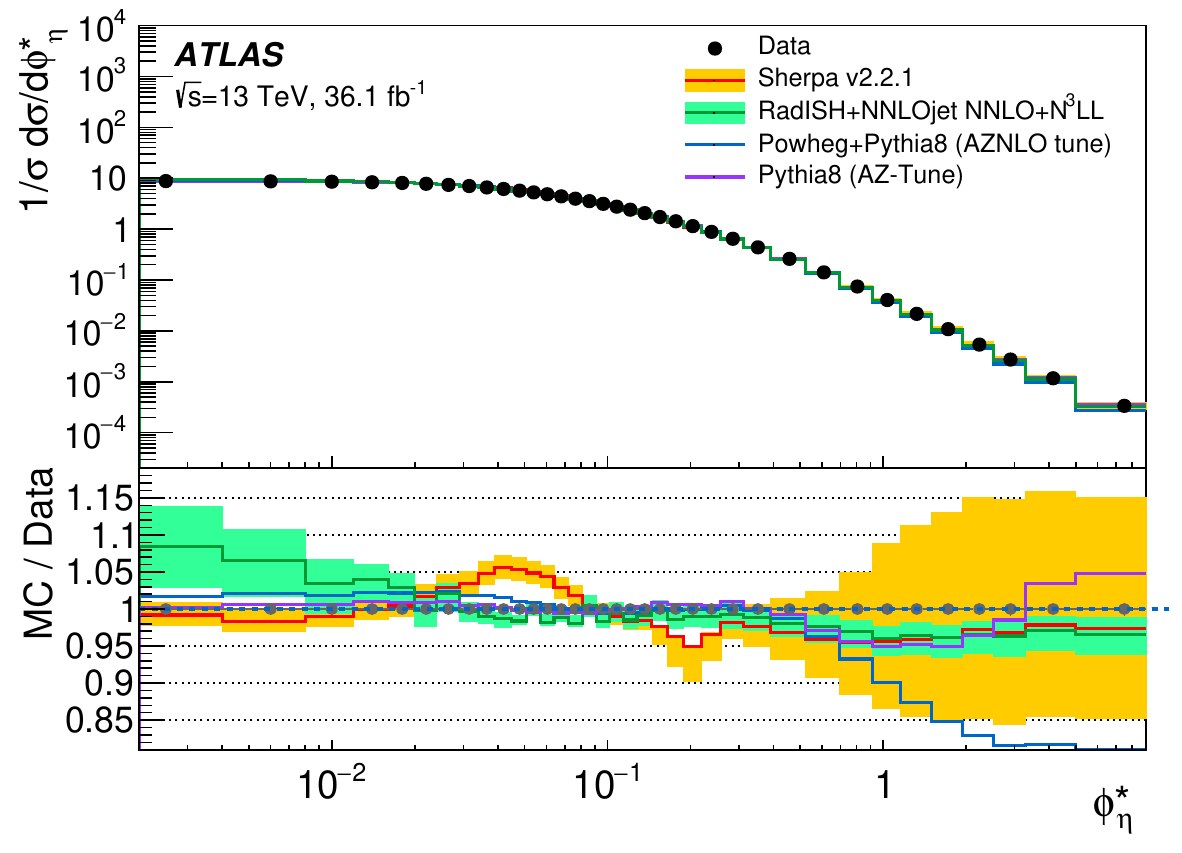}
\caption{\label{fig:comparison} Comparison of the normalized \Zpt{}
(left) and \phiStar{} (right) distributions predicted by different
computations: \PYTHIA{}8 with the AZ tune, \POWHEG{}+\PYTHIA{}8 with
the AZNLO tune, \SHERPA v2.2.1 and \textsc{RadISH}
with the Born level combined measurement. The uncertainties of the
measurement are shown as vertical bars and uncertainties of the
\SHERPA and \textsc{RadISH} predictions are indicated by the
coloured bands.}
\end{center}
\end{figure}
 
The differential measurements are compared with a variety of predictions of the
\Zpt{} and $\phiStar$ spectra that are based on different theoretical approaches to take
into account both the soft and hard emissions from the initial state (ISR).
Unless stated otherwise, the predictions do not consider NLO EW effects.
The comparisons between
the combined result corrected to QED Born level and the various
predictions are shown in Figures~\ref{fig:comparison} and \ref{fig:comparison2}. Systematic
uncertainties in the theoretical predictions are evaluated for
this comparison where feasible.
 
The first prediction is obtained from \PYTHIA{}8 with
matrix elements at LO in $\alphas$ supplemented with a parton shower
with the AZ set of tuned parameters~\cite{STDM-2012-23}. The AZ tune
optimized the intrinsic $k_\mathrm{T}$ and parton shower ISR
parameters to optimally describe the ATLAS 7~\TeV{} \Zpt{} and \phiStar{}
data~\cite{STDM-2012-23, STDM-2012-06}. It was later used in the
measurement of the $W$-boson mass using 7~\TeV{}
data~\cite{STDM-2014-18}, which requires a high-precision description
of the $W$-boson transverse momentum spectrum at low \pT.

The second prediction is based on \POWHEG{}+\PYTHIA{}8 using NLO matrix elements
with the \PYTHIA{}8 parton shower parameters set according to
the AZNLO tune~\cite{STDM-2012-23} derived using the same data as the
\PYTHIA{}8 AZ tune. The predictions using the AZ and AZNLO tunes
describe the 13~\TeV{} data to within 2--4\% in the region of low $\Zpt <
40$~\GeV{} and $\phiStar < 0.5$, and in this region the prediction using the
\PYTHIA{}8 AZ tune is the one that agrees best with the data. This
shows that predictions based on tunes to 7~\TeV{} collision data can also
provide a good description at significantly higher centre-of-mass
energies for low \Zpt{}. At high \Zpt{} these predictions are well below the data due to missing
higher-order matrix elements, similar to the situation observed at lower $\sqrt{s}$.
 
The third prediction is simulated with the
\SHERPA~v2.2.1~\cite{Bothmann:2019yzt} generator. In this set-up,
NLO-accurate matrix elements for up to two partons, and LO-accurate matrix
elements for up to four partons are calculated with the
Comix~\cite{Gleisberg:2008fv} and
OpenLoops~\cite{Cascioli:2011va,Denner:2016kdg} libraries. The default
\SHERPA parton shower~\cite{Schumann:2007mg} based on Catani--Seymour
dipole factorisation and the cluster hadronization model~\cite{Winter:2003tt} is
used with the dedicated set of tuned parameters developed by the
authors for the NNPDF3.0nnlo PDF set~\cite{Ball:2014uwa}. The NLO
matrix elements of a given parton multiplicity are matched to the parton
shower using a colour-exact variant of the MC@NLO
algorithm~\cite{Hoeche:2011fd}. Different parton multiplicities are then
merged into an inclusive sample using an improved CKKW matching
procedure~\cite{Catani:2001cc,Hoeche:2009rj} which is extended to NLO
accuracy using the MEPS@NLO prescription~\cite{Hoeche:2012yf}. The
merging threshold is set to $20$~\GeV. Uncertainties from missing higher
orders are evaluated~\cite{Bothmann:2016nao} using seven variations of
the QCD factorization and renormalization scales in the matrix
elements by factors of $0.5$ and $2$, avoiding variations in opposite
directions. For the computation of uncertainties in the normalized
spectra the effect of a certain variation is fully
correlated across the full spectrum and an envelope of all variations
is taken at the end, which results in uncertainties of 3--4\% at low
\Zpt{} and up to 25\% at high \Zpt{}. The effects of uncertainties in the PDF set are
evaluated using 100 replica variations and are found to be very small,
typically $<1\%$ up to $\Zpt < 100$~\GeV{} and a few percent
above. \SHERPA does describe the data in the high $\Zpt > 30$~\GeV{}
and $\phiStar > 0.1$ region to within about 4\% up to the point where
statistical uncertainties in the data exceed that level, which is
better than the uncertainty estimate obtained
from scale variations. On the other hand, the \SHERPA prediction
disagrees with the shape of the data at low $\Zpt < 25$~\GeV{} and
somewhat less with the \phiStar{} distribution. The data may be useful
in improving the parton shower settings in this regime.
 
Finally a prediction based on the
\textsc{RadISH} program \cite{Bizon:2019zgf, Bizon:2018foh} is
presented that combines a fixed-order NNLO prediction of
$Z$+jet production ($O(\alphas^3)$) from
\textsc{NNLOjet}~\cite{Gehrmann-DeRidder:2016jns} with resummation of
$\log(\Mll/\Zpt)$ terms at next-to-next-to-next-to-leading-logarithm (N$^3$LL)
accuracy~\cite{Bizon:2017rah}. The NNPDF3.1nnlo set of
PDFs~\cite{Ball:2017nwa} is used with QCD
scales set to
$\muR=\muF=\sqrt{(\Mll)^2+(\Zpt)^2}$ and the resummation scale set to $Q = \Mll/2$. Uncertainties in this prediction are derived from
variations of $\muR$ and $\muF$ in the same way as for the \SHERPA
prediction described above and, in addition, two variations of $Q$ by a
factor of two up and down, assuming that the effects of scale variations
are fully correlated across the full spectrum. Within the
uncertainties of typically 1--3\% the \textsc{RadISH} prediction
agrees with the data over the full spectrum of \Zpt{} and \phiStar{}, apart from a small tension in the very low \Zpt{} and \phiStar{} region where non-perturbative effects are relevant, highlighting
the benefits of this state-of-the-art prediction.
 
\begin{figure}[t]
\begin{center}
\includegraphics[width=0.495\textwidth]{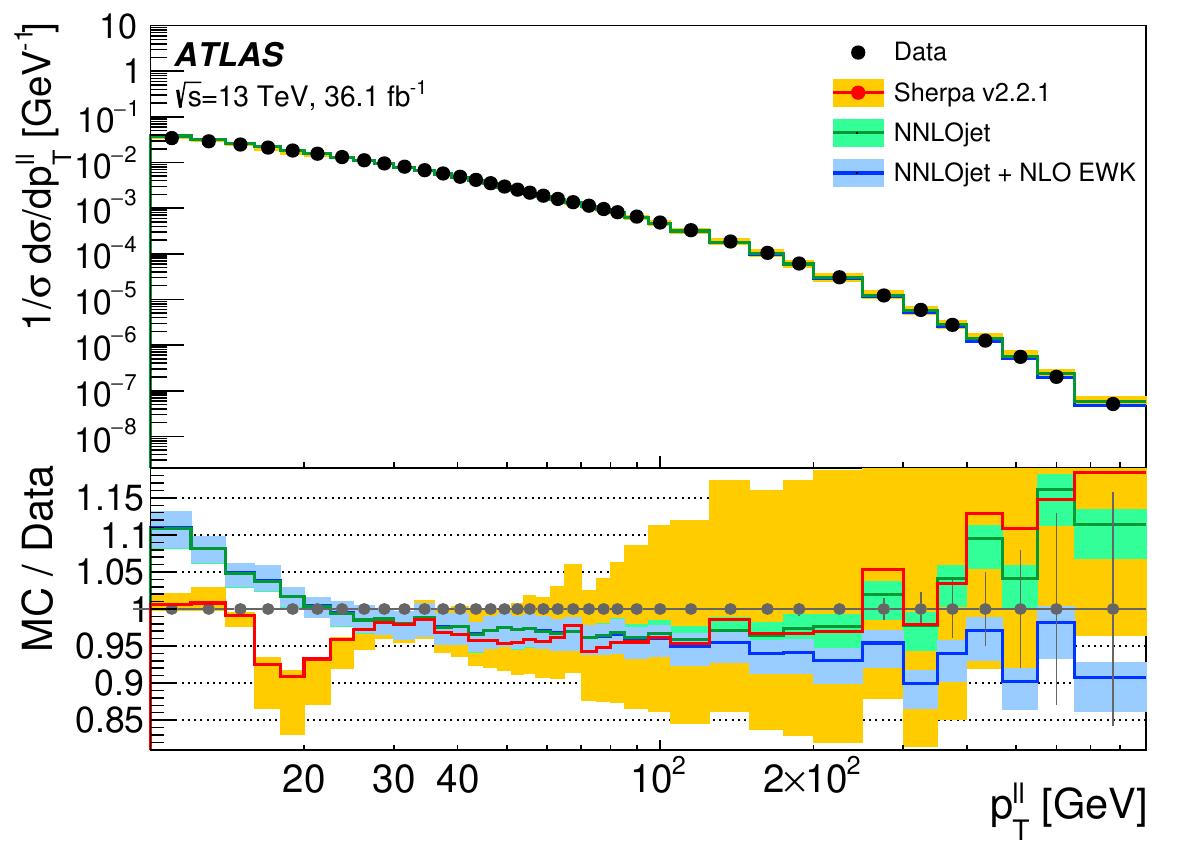}
\caption{\label{fig:comparison2} Comparison of the normalized \Zpt{}
distribution in the range $\Zpt > 10$~\GeV. The Born level combined
measurement is compared with predictions by \SHERPA v2.2.1,
fixed-order \textsc{NNLOjet} and \textsc{NNLOjet} supplied with NLO
electroweak corrections. The uncertainties in the measurement are
shown as vertical bars and the uncertainties in the predictions
are indicated by the coloured bands.}
\end{center}
\end{figure}
 
Figure \ref{fig:comparison2} compares the \Zpt{} measurement with predictions
in the range of $\Zpt > 10$~\GeV. In addition to the \SHERPA
prediction described above, the data are compared with the fixed-order
\textsc{NNLOjet} prediction described above both with and without NLO
EW corrections~\cite{Denner:2011vu}. The \textsc{NNLOjet} prediction
is only expected to describe the data at sufficiently large $\Zpt{}
\gtrsim 15\,\GeV$, while deviations for smaller values are expected
due to large logarithms
$\ln(\Mll/\Zpt)$~\cite{Gehrmann-DeRidder:2016jns}.  At the highest
\Zpt{} values probed, the application of NLO EW leads to a suppression
of up to $20\%$ due to large Sudakov logarithms. The theoretical
uncertainties on these corrections are not shown, but have been
elsewhere estimated to be up to 5\% for $\Zpt{} \approx
1\,\TeV$~\cite{Lindert:2017olm}. In this region, \textsc{NNLOjet}
without NLO EW corrections is generally above the data, and when
including these corrections it tends to be lower than the
data. However, the difference is not significantly larger than the
uncertainties in the measurements.
% End of text imported from the .//Sec_5_Results.tex input file
% The next lines are included from the .//Sec_6_Conclusion.tex input file
\section{Conclusion} 
 
Measurements of the \Zgee and \Zgmm cross-sections, differential in
the transverse momentum and \phiStar{}, have been performed in a
fiducial volume defined by $\pt^\ell > 27$~\GeV, $|\eta_\ell| < 2.5$ and
$66 < \Mll < 116$~\GeV, using $36.1~\ifb$ of data from
proton--proton collisions recorded in 2015 and 2016 at a
centre-of-mass energy of 13~\TeV{} with the ATLAS experiment at the
LHC. This data-set allows coverage of a kinematic range up to the $\TeV$-range. The cross-section results from the individual channels were combined and
good agreement between the two was observed. The relative precision of
the combined result is better than 0.2\% for $\Zpt{}<30\,\GeV$, which
provides crucial information to validate and tune MC event generators
and will constrain models of vector-boson production in future
measurements of the $W$-boson mass. 
 
The integrated fiducial cross-section measurements are compared with fixed-order perturbative QCD predictions. Differential spectra in \Zpt{} and
\phiStar{} are compared with a selection of calculations implementing
resummation and non-perturbative effects through parton showers or
analytic calculations. The predictions based on the \PYTHIA{}8 parton
shower with parameters tuned to 7 \TeV{} data are found to describe
the 13 \TeV{} data well at low \Zpt{} and \phiStar{}. The \SHERPA
prediction based on merging of higher-order, high-multiplicity matrix
elements gives an excellent description of the data at high $\Zpt$,
while the very accurate \textsc{RadISH} NNLO+N$^3$LL prediction agrees
with data for the full spectrum. The fixed-order \textsc{NNLOjet}
prediction with and without NLO EW effects describes the data well for high $\Zpt$.
% End of text imported from the .//Sec_6_Conclusion.tex input file
 
\clearpage
\section*{Acknowledgements}
% The next lines are included from the .//acknowledgements/Acknowledgements.tex input file

We thank CERN for the very successful operation of the LHC, as well as the
support staff from our institutions without whom ATLAS could not be
operated efficiently.
 
We acknowledge the support of ANPCyT, Argentina; YerPhI, Armenia; ARC, Australia; BMWFW and FWF, Austria; ANAS, Azerbaijan; SSTC, Belarus; CNPq and FAPESP, Brazil; NSERC, NRC and CFI, Canada; CERN; ANID, Chile; CAS, MOST and NSFC, China; COLCIENCIAS, Colombia; MSMT CR, MPO CR and VSC CR, Czech Republic; DNRF and DNSRC, Denmark; IN2P3-CNRS and CEA-DRF/IRFU, France; SRNSFG, Georgia; BMBF, HGF and MPG, Germany; GSRT, Greece; RGC and Hong Kong SAR, China; ISF and Benoziyo Center, Israel; INFN, Italy; MEXT and JSPS, Japan; CNRST, Morocco; NWO, Netherlands; RCN, Norway; MNiSW and NCN, Poland; FCT, Portugal; MNE/IFA, Romania; JINR; MES of Russia and NRC KI, Russian Federation; MESTD, Serbia; MSSR, Slovakia; ARRS and MIZ\v{S}, Slovenia; DST/NRF, South Africa; MICINN, Spain; SRC and Wallenberg Foundation, Sweden; SERI, SNSF and Cantons of Bern and Geneva, Switzerland; MOST, Taiwan; TAEK, Turkey; STFC, United Kingdom; DOE and NSF, United States of America. In addition, individual groups and members have received support from BCKDF, CANARIE, Compute Canada, CRC and IVADO, Canada; Beijing Municipal Science \& Technology Commission, China; COST, ERC, ERDF, Horizon 2020 and Marie Sk{\l}odowska-Curie Actions, European Union; Investissements d'Avenir Labex, Investissements d'Avenir Idex and ANR, France; DFG and AvH Foundation, Germany; Herakleitos, Thales and Aristeia programmes co-financed by EU-ESF and the Greek NSRF, Greece; BSF-NSF and GIF, Israel; La Caixa Banking Foundation, CERCA Programme Generalitat de Catalunya and PROMETEO and GenT Programmes Generalitat Valenciana, Spain; G\"{o}ran Gustafssons Stiftelse, Sweden; The Royal Society and Leverhulme Trust, United Kingdom.
 
The crucial computing support from all WLCG partners is acknowledged gratefully, in particular from CERN, the ATLAS Tier-1 facilities at TRIUMF (Canada), NDGF (Denmark, Norway, Sweden), CC-IN2P3 (France), KIT/GridKA (Germany), INFN-CNAF (Italy), NL-T1 (Netherlands), PIC (Spain), ASGC (Taiwan), RAL (UK) and BNL (USA), the Tier-2 facilities worldwide and large non-WLCG resource providers. Major contributors of computing resources are listed in Ref.~\cite{ATL-SOFT-PUB-2020-001}.
 
% End of text imported from the .//acknowledgements/Acknowledgements.tex input file

\printbibliography

\clearpage \input{atlas_authlist}
\end{document}

%% file: atlas_authlist.tex
% ATLAS Collaboration author list
% Reference date of STDM-2018-14 is 2019-08-05
% Author list last updated on date 26-FEB-20
% Data extracted on 26-Feb-2020 for paper reference STDM-2018-14
% at 3:54pm
 
\begin{flushleft}
{\Large The ATLAS Collaboration}

\bigskip

G.~Aad$^\textrm{\scriptsize 102}$,    
B.~Abbott$^\textrm{\scriptsize 129}$,    
D.C.~Abbott$^\textrm{\scriptsize 103}$,    
A.~Abed~Abud$^\textrm{\scriptsize 36}$,    
K.~Abeling$^\textrm{\scriptsize 53}$,    
D.K.~Abhayasinghe$^\textrm{\scriptsize 94}$,    
S.H.~Abidi$^\textrm{\scriptsize 167}$,    
O.S.~AbouZeid$^\textrm{\scriptsize 40}$,    
N.L.~Abraham$^\textrm{\scriptsize 156}$,    
H.~Abramowicz$^\textrm{\scriptsize 161}$
% ...
\bigskip
\\

% ...
$^{36}$CERN, Geneva; Switzerland.\\
% ...
$^{40}$Niels Bohr Institute, University of Copenhagen, Copenhagen; Denmark.\\
% ...
$^{53}$II. Physikalisches Institut, Georg-August-Universit\"{a}t G\"ottingen, G\"ottingen; Germany.\\
% ...
$^{94}$Department of Physics, Royal Holloway University of London, Egham; United Kingdom.\\
% ...
$^{102}$CPPM, Aix-Marseille Universit\'e, CNRS/IN2P3, Marseille; France.\\
$^{103}$Department of Physics, University of Massachusetts, Amherst MA; United States of America.\\
% ...
$^{129}$Homer L. Dodge Department of Physics and Astronomy, University of Oklahoma, Norman OK; United States of America.\\
% ...
$^{156}$Department of Physics and Astronomy, University of Sussex, Brighton; United Kingdom.\\
% ...
$^{161}$Raymond and Beverly Sackler School of Physics and Astronomy, Tel Aviv University, Tel Aviv; Israel.\\
% ...
$^{167}$Department of Physics, University of Toronto, Toronto ON; Canada.\\
% ...

\end{flushleft}

% Created with Glance <Atlas.Glance@cern.ch>